\documentclass[11pt]{article}
\usepackage{cite}
\usepackage{hyperref}
\usepackage{graphicx}
\usepackage{amsmath}
\usepackage{amssymb}
\usepackage{color}
\usepackage{braket}
\usepackage[]{epsfig}
\usepackage{enumitem}
\usepackage{comment}

\newcommand\snowmass{\begin{center}\rule[-0.2in]{\hsize}{0.01in}\\\rule{\hsize}{0.01in}\\
\vskip 0.1in Submitted to the  Proceedings of the US Community Study\\ 
on the Future of Particle Physics (Snowmass 2021)\\ 
\rule{\hsize}{0.01in}\\\rule[+0.2in]{\hsize}{0.01in} \end{center}}

\begin{document}

\snowmass

%\begin{center} \begin{Large} \emph{
%White Paper for Snowmass 2021
%}\end{Large} \end{center}

\begin{center} \begin{LARGE} \textbf{
Physics in the $\tau$-charm Region at BESIII
} \end{LARGE} \end{center}

\begin{flushleft}
\vspace{0.1in}
\small{ \emph{BESIII Spokespersons:} }
H.~B.~Li$^{1,58}$,
W.~Gradl$^{31}$,
X.~R.~Lyu$^{58}$,
C.~Z.~Yuan$^{1,58}$ \\
\vspace{0.1in}
\small{ \emph{Executive Board:} }
M.~Y.~Dong$^{1,53,58}$,
Y.~N.~Gao$^{42,g}$,
X.~C.~Lou$^{1,53,58}$,
M.~Maggiora$^{69A,69C}$,
Y.~J.~Mao$^{42,g}$,
R.~E.~Mitchell$^{25}$,
H.~P.~Peng$^{66,53}$,
X.~Y.~Shen$^{1,58}$ \\
\vspace{0.1in}
\small{ \emph{Physics Coordinators:} }
L.~Y.~Dong$^{1,58}$,
D.~Y.~Wang$^{42,g}$,
B.~Zheng$^{67}$ \\
\vspace{0.1in}
\small{ \emph{Light Hadron Physics Contacts:} }
S.~S.~Fang$^{1,58}$,
I.~Garzia$^{27A,27B}$,
N~H\"usken$^{25,31}$,
A.~Kupsc$^{40,70}$,
M.~Pelizaeus$^{4}$ \\
\vspace{0.1in}
\small{ \emph{Charmonium Physics Contacts:} }
Y.~P.~Guo$^{10,f}$,
W.~K\"uhn$^{33}$,
K.~Zhu$^{1}$ \\
\vspace{0.1in}
\small{ \emph{Charm Physics Contacts:} }
B.~C.~Ke$^{75}$,
J.~Libby$^{24}$,
H.~L.~Ma$^{1}$,
G.~Wilkinson$^{64}$ \\
\vspace{0.1in}
\small{ \emph{R Values, QCD, and tau Physics Contacts:} }
R.~Aliberti$^{31}$,
C.~F.~Redmer$^{31}$,
J.~Y.~Zhang$^{1}$,
X.~R.~Zhou$^{66,53}$ \\
\vspace{0.1in}
\small{ \emph{Exotic Decays and New Physics Contacts:} }
Z.~Q.~Liu$^{45}$,
D.~Y.~Wang$^{42,g}$,
Z.~Y.~You$^{54}$,
M.~G.~Zhao$^{39}$ \\
\vspace{0.1in}
\small{ \emph{Primary White Paper Contact:} }
R.~E.~Mitchell$^{25}$ \texttt{<remitche@indiana.edu>} \\
\end{flushleft}

\begin{scriptsize}
\begin{center}
$^{1}$ Institute of High Energy Physics, Beijing 100049, People's Republic of China\\
$^{4}$ Bochum Ruhr-University, D-44780 Bochum, Germany\\
$^{10}$ Fudan University, Shanghai 200433, People's Republic of China\\
$^{24}$ Indian Institute of Technology Madras, Chennai 600036, India\\
$^{25}$ Indiana University, Bloomington, Indiana 47405, USA\\
$^{27}$ INFN Sezione di Ferrara, (A)INFN Sezione di Ferrara, I-44122, Ferrara, Italy; (B)University of Ferrara, I-44122, Ferrara, Italy\\
$^{31}$ Johannes Gutenberg University of Mainz, Johann-Joachim-Becher-Weg 45, D-55099 Mainz, Germany\\
$^{33}$ Justus-Liebig-Universitaet Giessen, II. Physikalisches Institut, Heinrich-Buff-Ring 16, D-35392 Giessen, Germany\\
$^{39}$ Nankai University, Tianjin 300071, People's Republic of China\\
$^{40}$ National Centre for Nuclear Research, Warsaw 02-093, Poland\\
$^{42}$ Peking University, Beijing 100871, People's Republic of China\\
$^{45}$ Shandong University, Jinan 250100, People's Republic of China\\
$^{53}$ State Key Laboratory of Particle Detection and Electronics, Beijing 100049, Hefei 230026, People's Republic of China\\
$^{54}$ Sun Yat-Sen University, Guangzhou 510275, People's Republic of China\\
$^{58}$ University of Chinese Academy of Sciences, Beijing 100049, People's Republic of China\\
$^{64}$ University of Oxford, Keble Road, Oxford OX13RH, United Kingdom\\
$^{66}$ University of Science and Technology of China, Hefei 230026, People's Republic of China\\
$^{67}$ University of South China, Hengyang 421001, People's Republic of China\\
$^{69}$ University of Turin and INFN, (A)University of Turin, I-10125, Turin, Italy; 
%(B)University of Eastern Piedmont, I-15121, Alessandria, Italy; 
(C)INFN, I-10125, Turin, Italy\\
$^{70}$ Uppsala University, Box 516, SE-75120 Uppsala, Sweden\\
$^{75}$ Zhengzhou University, Zhengzhou 450001, People's Republic of China\\
$^{f}$ Also at Key Laboratory of Nuclear Physics and Ion-beam Application (MOE) and Institute of Modern Physics, Fudan University, Shanghai 200443, People's Republic of China\\
$^{g}$ Also at State Key Laboratory of Nuclear Physics and Technology, Peking University, Beijing 100871, People's Republic of China\\
\end{center}
\end{scriptsize}

\newpage

% RM EDITED: 4/14
\abstract{
The Beijing Spectrometer (BESIII) collaboration uses $e^+e^-$ collisions in the $\tau$-charm energy region 
to study a broad spectrum of topics.  These include studies of light mesons and light baryons, studies of charmonium, including exotic mesons and baryons containing charmonium, studies of charmed mesons and baryons, studies of QCD and $\tau$ physics, as well as searches for new physics.  The following is a Snowmass white paper that outlines the BESIII accomplishments and potential in each of these areas.
}

\tableofcontents

\newpage

% RM EDITED: 4/18
\section{Introduction to the BESIII Experiment}

The BESIII collaboration,
which operates the BESIII spectrometer at the Beijing Electron Positron Collider~(BEPCII), 
uses $e^+e^-$ collisions with center-of-mass (CM) energies ranging from 2.0 to 5.0~GeV
to study the broad spectrum of physics accessible in the tau-charm energy region.
Since the start of operations in 2009, BESIII has collected more than 35~fb$^{-1}$ of data, comprising several world-leading data samples, including: 
\begin{itemize}
\item 10~billion $J/\psi$ decays, giving unprecedented access to the light hadron spectrum;
\item 2.7~billion $\psi(2S)$ decays, allowing precision studies of charmonium and its transitions; 
 \item targeted data samples above 4~GeV, providing unique access to exotic XYZ hadrons;
\item 3~fb$^{-1}$ of data at the $\psi(3770)$ mass, providing a large sample of $D$ decays and quantum-correlated $D^{0}\bar{D}^{0}$ pairs, crucial for global flavor physics efforts; 
\item 3~fb$^{-1}$ at 4.18~GeV, near the peak of the $D_s^{\pm} D_s^{*\mp}$ cross section, for $D_s$ studies; 
\item more than 3~fb$^{-1}$ above $\Lambda_c\bar{\Lambda}_c$ threshold for precision $\Lambda_c$ studies; and
\item fine-scan samples for measurements of $R$, the mass of the $\tau$, and electromagnetic form factors.
\end{itemize}
The program will continue for at least the next 5-10 years, building on the data sets already collected, and ensuring the BESIII collaboration will remain a key player in future global efforts in hadron spectroscopy, flavor physics, and searches for new physics.  The maximum energy of BEPCII will soon be upgraded to 5.6~GeV, and there are plans to more than double the BEPCII luminosity at high CM energies by increasing the maximum achievable beam currents.   
Below we briefly outline a few highlights from BESIII, how these achievements have contributed to global physics efforts, and how the next era at BESIII will build on this momentum.  More details and references can be found in a recent white paper describing the future physics program at BESIII~\cite{whitepaper}.

% RM EDITED: 4/14
\section{The BEPCII-U Upgrade}

BEPCII delivered its first physics data in 2009 on the $\psi(2S)$ resonance. Since then,
BESIII has collected more than 35~fb$^{-1}$ of integrated luminosity at different CM energies from 2.0 to 4.94 GeV. In order to extend the physics potential of BESIII,  two upgrade plans for BEPCII were proposed and approved in 2020. The first upgrade will increase the maximum beam energy to~2.8 GeV (corresponding to a CM energy of 5.6~GeV), which will expand the energy reach of the collider into new territory.
The second upgrade will increase the peak luminosity by a factor of 3 for beam energies from 2.0 to 2.8~GeV (CM energies from 4.0 to 5.6~GeV). 

To perform these upgrades, we will increase the beam current and suppress bunch lengthening, 
which will require higher RF voltage.  
The RF,
cryogenic, and feedback systems will be upgraded accordingly. 
Nearly all of the photon absorbers along the ring and some vacuum chambers will also be replaced in order to protect the machine from the heat of synchrotron radiation. 
The budget
is estimated to be about 200 million CNY and it will take about 3 years to prepare the upgraded components and half a year for installation and commissioning, which will start in June 2024 and finish in December 2024.  With these upgrades, 
BESIII will enhance its capabilities to explore XYZ physics and will have the unique ability to perform precision measurements of the production and decays of charmed mesons and baryons at threshold.

% RM EDITED: 4/14
\section{Light Hadron Physics}

% this is the Letter-of-Intent, but somehow makes for a nice introduction as well?
The centerpiece of the BESIII light hadron physics program is the recently-collected sample of 10~billion $J/\psi$ decays.  Radiative decays of the $J/\psi$ provide a gluon-rich environment ideal for the production of scalar, pseudoscalar, and tensor glueballs.  The combination of advancements in theory, especially lattice QCD, and continuing innovation in amplitude analysis techniques, such as coupled-channel analyses, have brought the study of glueballs into finer focus.  The $J/\psi$ data set also provides hundreds of other exclusive decay channels with well-defined initial and final states in which both the light quark meson and baryon spectra can be investigated in fine detail.  Through the decays $J/\psi\to\phi\eta^{(\prime)}$ and $J/\psi\to\gamma\eta^{(\prime)}$, BESIII has also gathered competitive samples of $\eta$ and $\eta^{\prime}$ decays, crucial for precision tests of chiral perturbation theory.

% RM EDITED: 4/14
\subsection{Light meson spectroscopy}
% to be done by Nils
In the naive quark model, mesons are described as bound states of a quark and an anti-quark. However, QCD allows for a much richer meson spectrum, including tetraquark states, mesonic molecules, hybrid mesons and glueballs. Lattice QCD predicts the lightest glueballs to be scalar, tensor and pseudo-scalar, allowing mixing with the conventional mesons of the same quantum numbers. Generally, glueballs are expected to be produced in gluon-rich processes. Radiative $J/\psi$ decays provide such an environment, so that the high-statistics $J/\psi$ sample puts us in a unique position to study glueball candidates. Partial wave analyses of the radiative decays $J/\psi\to\gamma\pi^0\pi^0$ \cite{gampipi}, $\gamma K_S^0 K_S^0$ \cite{gamKsKs} 
and $\gamma\eta\eta$ \cite{gametaeta} reveal a strong production of the $f_0(1710)$ and $f_0(2100)$.
One might speculate that these resonances have a large gluonic component. Similarly, the tensor meson $f_2(2340)$ is strongly produced in the radiative decays $J/\psi\to\gamma\eta\eta$ and $\gamma\phi\phi$ \cite{gamphiphi}, rendering it a good candidate for a tensor glueball. Two recent coupled channel analyses \cite{bonn,jpac} of our data on radiative $J/\psi$ decays came to different conclusions concerning the number of contributing resonances and the identification of a glueball candidate, so that additional studies using the full 10~billion $J/\psi$ data sample will be of high importance in the future.

In the search for the pseudo-scalar glueball, the decay $J/\psi\to\gamma\eta^\prime\pi^+\pi^-$ has proven to be particularly interesting \cite{X1835}. Here, the $X(1835)$ can be observed with a lineshape that appears to be distorted at the proton anti-proton threshold, indicating a potential $p\bar{p}$ bound-state or resonance. In addition, the higher mass structures $X(2120)$ and $X(2370)$ are observed, although their spin-parity remains to be determined, a task that will be possible using the new, high precision $J/\psi$ data.

Motivated by multiple studies of the hybrid meson candidate $\pi_1(1600)$, which among many other production mechanisms is found in the decay $\chi_{c1}\to \pi_1(1600)^\pm\pi^\mp \to \eta^\prime\pi^+\pi^-$ \cite{pi1chic}, a recent search for the isoscalar partner states $\eta_1$ and $\eta_1^\prime$ in the radiative decays $J/\psi\to\gamma \eta \eta^\prime$ revealed a significant contribution from a new structure $\eta_1(1855)$ with exotic quantum numbers $J^{PC}=1^{-+}$ \cite{eta11,eta12}. While it is too early to say whether the $\eta_1(1855)$ is indeed an isoscalar hybrid meson, future studies of alternative decay modes will help reveal its nature.

The light scalar mesons $f_0(980)$ and $a_0(980)$ are frequently discussed as potential multiquark candidates, either as $K\bar{K}$ molecules or as compact tetraquark states. One possible way to probe their structure is the study of $f_0(980)$ - $a_0(980)$ mixing first observed by BESIII in the isospin-violating processes $J/\psi\to \phi a^0_0(980)$ and $\chi_{c1}\to \pi^0 f_0(980)$ \cite{mixing1,mixing2}. These results provide constraints in the development of theoretical models concerning the $f_0(980)$ and $a_0(980)$.

With 10~billion $J/\psi$ decays and the newly acquired 2.7~billion $\psi(2S)$, precision studies of conventional and exotic mesons, including multiquark states, glueballs and hybrid mesons, in radiative and hadronic $J/\psi$, $\psi(2S)$ and $\chi_{cJ}$ decays will be key tasks in the coming years.

% RM EDITED: 4/14
\subsection{Light meson decays}
% to be done by Isabella
% improve and check the references
Light meson decays play a fundamental role in the study of strong interactions in the non-perturbative region,
for the measurement of Standard Model~(SM) parameters, and in searches for new physics processes.
In particular, the $\eta$ and $\eta'$ mesons, the neutral members of the ground state pseudoscalar nonet, are important for understanding 
low energy quantum QCD~\cite{RevModPhys.91.015003}.
The 10~billion $J/\psi$ events collected at BESIII offer an unique opportunity to investigate all these aspects, as well as the
search for rare $\eta$ and $\eta'$ decays needed to test fundamental QCD symmetries and probe physics beyond the SM.
The decays of $J/\psi\rightarrow \gamma \eta(\eta')$ and $J/\psi\rightarrow \phi \eta(\eta')$
provide clean and efficient sources of $\eta/\eta'$ mesons for the decay studies.

The observation of new $\eta'$ decay modes, including $\eta'\rightarrow\rho^{\mp}\pi^{\pm}$~\cite{PhysRevLett.118.012001}, 
$\eta'\rightarrow\gamma e^+ e^-$~\cite{PhysRevD.92.012001},  and $\eta'\rightarrow 4\pi$~\cite{PhysRevLett.112.251801,PhysRevLett.113.039903}
 have been reported for the first time using about $10^9$ $J/\psi$ decays.
Using the same data set, the  branching fractions of the five dominant decay channels of the $\eta'$ were measured for the first time~\cite{PhysRevLett.122.142002}
using events in which the radiative photon converts to $e^+ e^-$. 
This is an effective technique 
for tagging inclusive decays of the $\eta'$ meson
and leads to a very high precision measurement of the absolute $\eta'$ branching fractions.
 
The double Dalitz decay $\eta'\rightarrow e^+ e^+ e^- e^-$
is of great interest for understanding the pseudoscalar transition form factor and the interaction between pseudoscalar and virtual photons.
It also could help in the reduction of the theoretical uncertainty of the muon anomalous magnetic moment thanks to the information
related to the meson-photon-photon vertices that contribute to the hadronic light-by-light term.
This process has not been observed to date, while the predicted branching fraction is of the order of $2\times10^{-6}$~\cite{Escribano_2018,petri2010anomalous}.
Another interesting study is the hadronic decay $\eta'\rightarrow \pi^0\pi^0\eta$
which is sensitive to the elastic $\pi\pi$ S-wave scattering lengths, and causes     
% which is sensitive to the elastic S-wave scattering of neutral pions charge-exchange rescattering $\pi^+\pi^- \rightarrow \pi^0\pi^0$, resulting to
a prominent cusp effect in the $\pi^0\pi^0$ invariant mass spectrum at the $\pi^+\pi^-$ mass threshold~\cite{refId0}.
The full $J/\psi$ data set collected by BESIII offers unique opportunities to investigate the cusp effect in this decay for which no evidence has yet been found.

Following the same analysis strategy of Ref.~\cite{PhysRevLett.122.142002},
the absolute branching fraction of the decay $J/\psi \rightarrow \gamma\eta$ has been measured with high precision using radiative photon conversions~\cite{PhysRevD.104.092004}, 
and the four dominant $\eta$ decays have been measured for the first time.
The $\eta/\eta'\rightarrow\gamma\pi^+\pi^-$ decay results are related to details of chiral dynamics; 
$\eta/\eta'\rightarrow 3\pi$ decays provide information on the up and down quark masses;
and the decay widths of $\eta/\eta'\rightarrow \gamma\gamma$ are related to the quark content of the two mesons.
Despite the impressive progress in the last years, many $\eta$ and $\eta'$ decays are still to be observed and explored.
%Thanks to the 10 billion of $J/\psi$ events collected at BESIII, the $\eta/\eta'$ events from radiative decays ($J/\psi\rightarrow \gamma \eta/\eta'$) and hadronic decays 
%($J/\psi\rightarrow \phi \eta/\eta'$ and $J/\psi\rightarrow \omega \eta/\eta'$ ) make possible more detailed studies end reach unprecedented precisions.
The full $J/\psi$ data set now available at BESIII makes possible more detailed studies with unprecedented precision.
It allows, in addition, an intensive investigation of the properties of the pseudoscalar states $\eta(1405)/\eta(1475)$~\cite{PhysRevLett.108.182001,PhysRevD.91.052017}; 
a thorough study of all states observed in the $1.4-1.5\,\rm{GeV}/c^2$ mass region;
a deep investigation of the $\omega\rightarrow\pi^+\pi^-\pi^0$ Dalitz plot~\cite{PhysRevD.98.112007}; 
and searches for rare $\omega$ decays.

% RM EDITED: 4/14
\subsection{Light baryon spectroscopy}
% to be done by Shuangshi

Although the quark model and lattice QCD have achieved significant success in the
interpretation of many static properties of baryons and the
excited resonances,  our present knowledge on baryon spectroscopy is
still in its infancy as many fundamental issues are still not well understood~\cite{Capstick:2000qj}.
One of  the outstanding problems  is that a substantial number of ‘missing $N^*$ states’ predicted by the quark model have not  yet been observed, a fact which has inspired both experimental and theoretical efforts during the last decade. 

The high production rate of baryons in charmonium decays, combined with  the large data samples of  $J/\psi$ and $\psi(2S)$ decays produced from $e^+e^-$ annihilations, provides excellent opportunities for studying excited baryons.  
Therefore, the BES experiment launched a program to study the excited baryon spectrum and a series of important achievements were made~\cite{BES:2001gvq,BES:2004gwe,BES:2009ufh,BESIII:2013xkm,BESIII:2012ssm,BESIII:2013bgg} using an effective
Lagrangian approach \cite{Benmerrouche:1996ij,Olsson:1976st}  and the extended
automatic Feynman Diagram Calculation (FDC) package\cite{Wang:1993dwa}. In addition to the well know excited nucleon states, a few new clear $N^*$ structures, the $N(2065)$, $N(2300)$, and $N(2570)$, were observed. These results also demonstrate that the BES experiment is a powerful instrument for investigating baryon spectroscopy. 

At present,  the search for hyperon resonances remains an important
challenge. Some of the lowest excitation resonances have not yet been
 experimentally resolved, which are necessary to establish the spectral pattern of hyperon resonances.  The large data samples of $J/\psi$ and $\psi(2S)$ decays accumulated by the BESIII experiment 
 enable us to complete the hyperon (e.g., 
$\Lambda^*$, $\Sigma^*$ and $\Xi^*$) spectrum and examine various
pictures for their internal structures.  Such pictures include a simple $3q$ quark structure or a more complicated structure with pentaquark components dominating.  In particular, $\psi(2S)$ decays, because of the larger mass of the $\psi(2S)$, have great potential to uncover new higher excitations of hyperons.

At BESIII, $10^{10}$ $J/\psi$ and $2.7\times 10^9$ $\psi(2S)$  decays are now
available, which offer great additional opportunities for investigating baryon spectroscopy. 
Together with other high-precision
experiments, such as  GlueX and JPARC, these very abundant and clean event samples
will bring the study
of baryon spectroscopy into a new era, and will
 make significant contributions to our understanding
of hadron physics in the non-perturbative regime.

% RM EDITED: 4/14
\subsection{Precision hyperon physics}
% to be done by Andrzej
Observation of a significant polarization of the  $\Lambda$ and $\bar\Lambda$ from $J/\psi\to\Lambda\bar\Lambda$ led to the revision of the decay asymmetry parameter $\alpha_\Lambda$~\cite{BESIII:2018cnd} and has shown BESIII has the potential to study properties of the ground-state (anti)hyperons. The branching fractions for
$J/\psi$ decays into a hyperon--antihyperon pair are relatively large, ${\cal O}(10^{-3})$, and thus the collected 10 billion $J/\psi$ decays can be used for precision studies of hyperon decays and tests of CP symmetry. The hyperon--antihyperon pair is produced in a well-defined spin-entangled state based on the two possible partial waves (parity symmetry in this strong decay allows for an $S$- and a $D$-wave). {The charge-conjugated decay modes of the hyperon and antihyperon can be measured simultaneously and their properties compared directly.} In the first round of analyses both the hyperon and antihyperon decay via  the common nonleptonic modes. The full data set will be used  to improve the precision of the CP-violation searches within these decays. The next stage will be to use a common decay of one of the (anti)hyperons to study rare decays of the produced partner. For example, the  kinematical constraints make it possible to perform complete reconstruction of the semileptonic decays of polarized hyperons. 

% RM EDITED: 4/14
\section{Charmonium and XYZ Physics}  
The well-established spectrum of charmonium states below open-charm threshold, including the $\eta_c(1S,2S)$, the $h_c(1P)$, and the $\chi_{cJ}(1P)$ states, can be cleanly accessed through $\psi(2S)$ hadronic and radiative transitions.  Large samples of $\psi(2S)$ decays and transitions, with a  sample of about 2.7~billion $\psi(2S)$, allow detailed studies of the behavior of QCD between the perturbative and non-perturbative regimes and are an ideal testing ground for effective field theories like NRQCD.  Above open-charm threshold, the charmonium spectrum has revealed exciting evidence for states beyond the simple $c\bar{c}$ model.  These XYZ states are produced at BESIII in ways complementary to the $B$-factories and the LHC.  
The $Y(4260)$, for example, was discovered by BaBar using initial state radiation; but BESIII, by producing the $Y(4260)$ directly, has revealed intriguing fine structure that is rewriting the story of the $Y$ states.  BESIII has also used hadronic decays of the $Y$ to discover manifestly exotic isovector $Z_c$ states and has used radiative transitions to study properties of the $X(3872)$. Furthermore, there are chances for BESIII to search for pentaquark states containing $c\bar{c}$ with data to be collected at CM energies above 5~GeV. Future luminosity and energy upgrades will push these studies into new territories.

% RM EDITED: 4/14
\subsection{Charmonium physics}

Below the open-charm threshold, the spin-triplet charmonium states are produced copiously in $e^+e^-$ annihilation and in $B$ decays so they are understood much better than the spin-singlet charmonium states, including the lowest lying S-wave state $\eta_c$, its radial excited partner $\eta_c(2S)$, and the P-wave spin-singlet state $h_c$. Since these three states can all be produced in $\psi(2S)$ decays, the 2.7 billion $\psi(2S)$ decays at BESIII make it possible to study their properties with improved precision. In addition, the unexpectedly large production cross section for $e^+e^-\to \pi^+\pi^-h_c$ in the BESIII high-energy region provides a new mechanism for studying the $h_c$ and $\eta_c$ (from $h_c\to \gamma \eta_c$). A few of these important physics topics include searching for additional decay modes of the $\eta_c$, $\eta_c(2S)$, and $h_c$ states, improving the precision of the masses and widths of the $\eta_c$, $\eta_c(2S)$ and $h_c$, and searching for hadronic and radiative transitions between the charmonium states. BESIII has contributed world best measurements of the properties of these states, see Ref.~\cite{whitepaper} form more details, and will improve these measurements in the future.
%The knowledge of the hadronic decay of $h_c$ and $\eta_c(2S)$ is limited, only six %channels are observed for $h_c$ and seven channels for $\eta_c(2S)$, many more %hadronic decay modes are expected and will be searched for at BESIII. 

Besides 2.7 billion $\psi(2S)$ events, two more data samples at $\sqrt{s}=3.65$~GeV and $3.682$~GeV have been collected, each with an integrated luminosity of approximately 400~pb${}^{-1}$. These samples can be used to study the interference between resonance decay and continuuum production and determine the hadronic decays of $\psi(2S)$ with better precision. This is  important to solve the long standing $\rho-\pi$ puzzle. Base on pQCD, the ratio of the branching fractions of $J/\psi$ and $\psi(2S)$ to the same final state is expected to hold to a reasonably good degree to $12\%$, and a similar relation is expected for the branching fractions of $\eta_c$ and $\eta_{c}(2S)$~\cite{Chao:1996sf}. Violations of these rules were reported~\cite{Mo:2006cy,Wang:2021dxw} and can be tested once more decay modes of the $\psi(2S)$ and $J/\psi$, and $\eta_{c}(2S)$ and $\eta_{c}$ states are observed. Furthermore, the 20~fb${}^{-1}$ data sample at the $\psi(3770)$ mass and scan samples around the peak can be used to study the non-$D\bar{D}$ decays of $\psi(3770)$.   

The search for new charmonium states has always been a high priority topic. Using data with CM energies above $4~\mathrm{GeV}$, it is possible to search for those unobserved states that are predicted by the potential models and study their properties once they are observed. These include the $1F$, $3S$, $2D$, $1G$, $3P$, $2F$, $4S$, $3D$, and $2G$ states. Among these, the $\psi_2(1D)$ state, also called the $\psi_2(3823)$, was confirmed at the BESIII experiment~\cite{BESIII:2015iqd} and the $\psi_3(1D)$
state was observed by the LHCb experiment (the $\psi_1(1S)$ state, the $\psi(3770)$, was observed many years ago). 

The excited vector charmonium states can be produced directly from $e^+e^-$ annihilation and can be studied in open-charm final states such as $D^{(*)}\bar{D}^{(*)}$, $D_s^{(*)}D_s^{(*)}$, %$D\bar{D}_J$, $D_s\bar{D}_{sJ}$, 
charmed baryon pair final states, hidden-charm final states, multi light hadron final states, and inclusive process. The cross sections of $e^+e^-\to D^{*+}{D}^{-}$ and $e^+e^-\to D^{*+}{D^{*-}}$ have been measured from $\sqrt{s}=4.085$ to $4.60~\mathrm{GeV}$ using $28$ data samples with typical precision of $3\%$ and $6\%$, respectively~\cite{BESIII:2021yvc}. Using current available data samples the precision of other two-body open-charm cross section are expected to be around $10\%$. The coupling of vector charmonium states to the open-charm meson pairs will provide crucial information in identifying the states in this region. The hadronic and radiative transitions between the (excited) charmonium states can be investigated to study the transition rates and decay dynamics. 
The cross section of $e^+e^-\to\eta J/\psi$~\cite{BESIII:2020bgb} shows an enhancement around the $\psi(4040)$ mass, while the cross sections of $e^+e^-\to\pi^+\pi^-\psi(3770)$~\cite{BESIII:2019tdo} and  $e^+e^-\to\pi^+\pi^-\psi_2(3823)$~\cite{BESIII:2022yga} show an enhancement 
around the $\psi(4415)$ mass. 
The process $e^+e^-\to\gamma\chi_{cJ}$ is studied to search for radiative transitions between the excited vector charmonium states and the $\chi_{cJ}$~\cite{BESIII:2021yal}. Whether are they produced via hadronic transitions from the excited vector charmonium states or via vector charmonium-like states is not yet clear and can be addressed using improved luminosity and more decay channels. 

Using the $e^+e^-\to\pi^+\pi^-\psi_2(3823)$ process, the most precise mass of the $\psi_2(3823)$ has been determined~\cite{BESIII:2022yga} and new decay modes of the $\psi_2(3823)$ have been searched for~\cite{BESIII:2021qmo}. These recent measurements at BESIII are examples that the transitions between charmonium states can also serve as production sources of non-vector charmonium states, and can be used to study the properties (mass, width and decay modes) of non-vector charmonium states. They will also be important study topics in the future at BESIII. 

%The search for new charmonium states has always been a high
%priority topic. With the data with CM energies above 4~GeV, it
%is possible to search for those unobserved states that are
%predicted by the potential models. These states include the
%excited P-wave spin-triplet states $\chi_{cJ}(2P)$ ($J=0$, $1$,
%$2$), the excited P-wave spin-singlet state $h_c(2P)$, the D-wave
%spin-triplet states $\psi_J(1D)$ ($J=2$, $3$; $J=1$ state, the
%$\psi(3770)$, was observed many years ago), and the D-wave
%spin-singlet state $\eta_{c2}(1D)$. Among these, the $\psi_2(1D)$
%state, also called the $\psi_2(3823)$, was discovered at BESIII,
%and the $\psi_3(1D)$ state was observed by LHCb.

%The spin-triplet charmonium states are produced copiously in $e^+e^-$
%annihilation and in $B$ decays so they are understood much better
%than the spin-singlet charmonium states, including the lowest
%lying S-wave state $\eta_c$, its radial excited partner
%$\eta_c(2S)$, and the P-wave spin-singlet state $h_c$. Since these
%three states can all be produced in $\psi(2S)$ decays, the world
%largest $\psi(2S)$ data sample at BESIII made it possible to study
%their properties with improved precision. In addition, the
%unexpected large production cross section of $e^+e^-\to \pi^+\pi^-h_c$ in
%BESIII energy region opened a new mechanism of studying $h_c$ and
%$\eta_c$ (from $h_c\to \gamma \eta_c$), and BESIII contributed
%world best measurements of the properties of these states and will
%be improve these measurements in the future.

% RM EDITED: 4/14
\subsection{Problems in XYZ physics}

The discovery of the XYZ states has subverted the ecosystem of the traditional studies of the charmonium
spectrum~\cite{Lebed:2016hpi,Chen:2016qju,Esposito:2016noz,Guo:2017jvc,Ali:2017jda,Olsen:2017bmm,Brambilla:2019esw,Yuan:2021wpg}. 
These exotic states cannot be embedded in the charm-anticharm frame that could be described successfully by a simple potential model, and must take novel configurations, such as tetraquarks, hybrids, or hadronic molecules. Studying them opens a new window for us to probe the nonperturbative QCD which underlies the formation of hadrons via strong interaction. BESIII has contributed to and will continue to uncover three kinds of broad problems of the XYZ states. These will be addressed in more detail as the ``Y problem'', ``Z problem'', and ``X problem'' below.

\paragraph{The $Y$ problem}.

BESIII has systematically measured the cross sections of various exclusive $e^+ e^-$ annihilations with hidden charm, open charm, and light hadronic final states. Some important results of these channels, such as $e^+e^- \to \pi^+\pi^- J/\psi$~\cite{BESIII:2016bnd}, $e^+e^- \to \pi^+\pi^- h_c$~\cite{BESIII:2016adj}, $e^+e^- \to \omega \chi_{c0}$~\cite{BESIII:2014rja}, $e^+ e^- \to \pi^+ D^0 D^{*-}$~\cite{BESIII:2018iea}, and $e^+ e^- \to \pi^+ \pi^- \psi(3682)$~\cite{BESIII:2017tqk,BESIII:2021njb}, show the lineshapes are complicated as a function of CM energy. Generally, BESIII's results have confirmed the $Y(4230)$, previously found by Belle, with hidden and open charm channels but shifted its mass to about $4220$ MeV, discovered a new structure named $Y(4390)$, and reproduced $Y(4660)$ with a hidden charm channel ($\pi^+ \pi^- \psi(2S)$) in addition to the $\Lambda_c$ pair. However, the extracted parameters of these $Y$ states are not consistent with each other in different channels. Furthermore, they deviate from the resonances observed in inclusive channels such as the $\psi(4040)$, $\psi(4160)$, and $\psi(4415)$ that are believed to be the conventional charmonia. This leads to the $Y$ problem. What are the exact lineshapes of these cross sections? Are these observed structures real new resonances? Or just results of some subtle kinematic effects? Will new patterns be discovered with new data? Will they be interpreted satisfactorily and match theoretical predictions? To solve the $Y$ problem, a detailed scan between 4.0 and 4.6~GeV is proposed~\cite{whitepaper}, with 500~pb$^{-1}$ per point, for points spaced at 10~MeV intervals. Till now, this target is partially achieved with about 22~fb$^{-1}$ integrated luminosity, and will be updated with larger maximum energy (5.6~GeV) after the upgrade of the BEPCII.

\paragraph{The $Z$ problem}.

The $Z_c(3900)$ was discovered at BESIII in the process $e^+e^- \to \pi^\mp Z_c^\pm$ with $Z_c^\pm \to \pi^\pm J/\psi$~\cite{BESIII:2013ris}, and the $Z_c(4020)$ was discovered in the process $e^+e^-\to \pi^\mp Z_c^\pm$ with $Z_c^\pm \to \pi^\pm h_c$~\cite{BESIII:2013ouc}. 
The $Z_c(3900)$ has also been observed in the open-charm channel $(D\bar{D}^* + c.c.)^\pm$~\cite{BESIII:2013qmu}, similarly the $Z_c(4020)$ was seen via the open-charm channel $(D^*\bar{D}^*)^\pm$~\cite{BESIII:2013mhi}.
Furthermore, neutral partners of these charged $Z_c$ states have been observed at BESIII via processes $e^+ e^- \to \pi^0 \pi^0 J/\psi$~\cite{BESIII:2015cld} and $e^+ e^- \to \pi^0 \pi^0 h_c$~\cite{BESIII:2014gnk}. 
BESIII has also determined the quantum numbers of the $Z_c(3900)$ to be $J^P = 1^+$~\cite{BESIII:2017bua}. 
Recently, BESIII has observed a new near-threshold structure in the $K^+$ recoil-mass spectra in $e^+ e^- \to K^+(D_s^- D^{*0}+ D_s^{*-}D^0)$~\cite{BESIII:2020qkh}. This structure, named $Z_{cs}(3985)$, is a good candidate of the charged hidden-charm tetraquark with strangeness. However, at the energy region higher than 4.3~GeV the data have revealed more complex structure in the Daliz plots of $e^+ e^- \to \pi^+ \pi^- J/\psi$. A similar situation is found in the $e^+e^- \to \pi^+\pi^-\psi(2S)$~\cite{BESIII:2017tqk}. This is the ``Z'' problem. Are the properties of these $Z_c$ states constant (corresponding to real resonant states) or energy dependent (corresponding to kinematic effects such as cusps or singularities)? 
What are the exact lineshapes of them? Can we find more decay patterns for them, especially for the newly discovered $Z_{cs}$ states? 
Are there spin multiplets of these $Z_{c}$ states? To answer these questions, BESIII may take advantage of the fine scan data mentioned before, but at a few points, a set of samples with very high statistics will be very helpful. BESIII currently has 1~fb$^{-1}$ of data for $e^+e^-$ cms energy at 4.23 and 4.42~GeV. Additional data including three or four points with an order of 5~fb$^{-1}$ or more per point is proposed to guarantee adequate statistics for amplitude analyses~\cite{whitepaper}. After the upgrade of BEPCII with triple the luminosity, this goal will be achieved more easily.

\paragraph{The $X$ problem}.

For the $X(3872)$, BESIII has discovered the process $e^+e^- \to \gamma X(3872)$~\cite{BESIII:2013fnz}, 
studied the open-charm decay and radiative transitions of the $X(3872)$~\cite{BESIII:2020nbj}, and has observed the hadronic transitions $X(3872) \to \pi^0 \chi_{c1}(1P)$~\cite{BESIII:2019esk} and $X(3872) \to \omega J/\psi$~\cite{BESIII:2019qvy}. 
The $X(3872)$, with its quantum numbers $J^{PC}=1^{++}$, has a mass very close to the predicted $\chi_{c1}(2P)$ state with a very narrow width. Then the ``X'' problem is can we separate the $X(3872)$ from the $\chi_{c1}(2P)$? Is the $X(3872)$ really exotic or conventional, or even a mixture state? 
Can we measure the line shape of the $X(3872)$? Are there other $X$ states (for example close to the $D^*\bar{D}^*$ threshold) that have not been observed yet? The related studies will benefit from the large scan and other data samples mentioned before. Furthermore, at $E_{cm}>4.7$ GeV with highly excited $\psi$ or $Y$ states produced, the hadronic transitions, that take larger production rates than the radiative transitions, are accessible. After the upgrade of BEPCII to its maximum CM energy, BESIII will have the ability to search for the $J^{++}$ states via hadronic transitions such as processes of $e^+ e^- \to \omega X$ and $e^+ e^- \to \phi X$.

\paragraph{Relationships}.

There are two kinds of relationships that deserve discussion. One is the relationship between XYZ states and conventional charmonia. For example, the $\chi_{c1}(2P)$ has a similar mass and the same $J^{PC}$ as the $X(3872)$. So a detailed understanding of the spectrum of the conventional $2P$ charmonium states, that include the spin triplet $\chi_{cJ}(2P)$ and singlet $h_c(2P)$, is crucial for understanding the nature of the $X(3872)$. This is also true for the other conventional charmonia and XYZ states under similar conditions. The studies of the conventional charmonia and exotic XYZ are complementary to each other. Understanding the relations between the two kinds of states, even the possible mixing between them, will be helpful for understanding the properties of the XYZ states. 
The other relationship is among the XYZ states. The analyses of processes $e^+e^- \to \gamma X(3872)$~\cite{BESIII:2013fnz} and $e^+ e^- \to \pi^0 \pi^0 J/\psi$~\cite{BESIII:2015cld} have already shown that there is evidence for the radiative transition $Y(4230)\to \gamma X(3872)$ and the hadronic transition $Y(4230) \to \pi^0 Z_c^0(3900)$. Searching for new transition modes and confirming these relations may be a unique chance for BESIII to reveal the nature of the internal structure of the XYZ states~\cite{Zhu:2021vtd}.

% RM EDITED: 4/14
\subsection{Pentaquark states}

The LHCb experiment reported the observation of three pentaquark states with a $c\bar{c}$ component in the $J/\psi p$ system via $\Lambda_b^0 \to J/\psi K^- p$~\cite{LHCb:2015yax,LHCb:2019kea}.
To confirm these states, further experimental research should be  pursued with the current available and the forthcoming experimental data~\cite{Shen:2015eua}. BESIII may search for such
and similar states with data to be collected at CM energies above 5~GeV in the processes $e^+e^- \to J/\psi p + X$, $\chi_{cJ} p + X$, $J/\psi \Lambda + X$, $\bar{D}^{(*)} p + X$, $D^{(*)} p + X$, and so on. It is clear that a systematic search for baryon-meson resonances should be pursed in various processes, where the baryon could be $p$, $\Lambda$, $\Sigma$, $\Sigma_c$, ..., and the meson could be $\eta_c$, $J/\psi$, $\chi_{cJ}$, $D^{(*)}$, etc. It is worth pointing out that the tetraquark and pentaquark candidates mentioned above have a pair of charm-anticharm quarks which may annihilate. Observations of states like $T_{cc}^+$
($cc\overline{ud}$) or $\mathrm{\Theta}_c^0$ ($uudd\bar{c}$) or $P_{cc}^0$ ($ccdd\bar{u}$) or similar serve as better evidence for multiquark states.

% RM ONLY LOOSELY EDITED

\section{Charm Physics}
BESIII has collected data samples with integrated luminosity of 4.4~fb$^{-1}$ at energies between 4.6 and 4.7 GeV, and will ultimately collect data samples with integrated luminosities of 20~fb$^{-1}$ at $\sqrt s=$ 3.773~GeV and 6~fb$^{-1}$ at $\sqrt s=$4.178~GeV; these CM energies are optimal values for the accumulation of  $\Lambda^{+}_{c}\bar{\Lambda}^{-}_c$, $D\bar{D}$, and $D^{*+}_{s}D_{s}^{-}$ events near threshold, respectively. The kinematics at threshold allow a double-tag technique to be employed where the full event can be reconstructed, even if it contains one undetected particle. 
%such as a neutrino $\nu$, $K_{L}^{0}$ meson or neutron. 
This provides a unique, low-background environment to measure the absolute branching fractions 
for charmed hadrons decaying to leptonic, semi-leptonic, and hadronic final states. Such measurements provide rigorous tests of QCD, CKM unitarity, and lepton flavor universality that complement similar studies of beauty hadrons. 
Furthermore, the 20~fb$^{-1}$ data sample of coherent $\psi(3770)\to D^{0}\bar{D}^{0}$ events allow measurements of the strong-phase differences between the $D^{0}$ and $\bar{D}^{0}$ that are essential inputs to determine the Unitarity Triangle angle $\gamma$ in an amplitude model-independent fashion from $B$ decays.
%at LHCb and its upgrades, as well as at Belle~II.
%These strong-phase differences can only be %determined using this data set to the required level of precision that would result in a systematic uncertainty below 0.5$^\circ$ for $\gamma$ measurements at future LHCb upgrades. 
These strong-phase measurements are also important ingredients of model-independent measurements of $D^{0}\bar{D}^{0}$ mixing and searches for indirect $\mathit{CP}$ violation in $D^{0}$ decay. 
For charmed baryons, in addition to precision branching fractions of the $\Lambda_c$. 
When the BEPCII CM energy is upgraded to reach 5.6~GeV, pairs of $\Sigma_c$ and $\Xi_c$ baryons will also be produced and studied. %Table~\ref{tab:prospect} presents the expected precision on some key %measurements of $D^{0(+)}$, $D^+_s$, and $\Lambda_c^+$ decays based on the %proposed data set, and a comparison with Belle II.
\subsection{Leptonic decays}
In the standard model, the partial widths of the leptonic decays
$D_{(s)}^+ \to \ell^+\nu_\ell$ can be written as
\begin{equation}
\Gamma(D_{(s)}^+ \to \ell^+\nu_\ell)=
     \frac{G^2_F} {8\pi}
      f^2_{D_{(s)}^+} |V_{cd(s)} |^2
      m^2_\ell m_{D_{(s)}^+}
    \left (1- \frac{m^2_\ell}{m^2_{D_{(s)}^+}}\right )^2,
\label{eq01}
\end{equation}
\noindent
where $G_F$ is the Fermi coupling constant,
$f_{D_{(s)}^+}$ is the $D_{(s)}^+$ decay constant,
$|V_{cd(s)}|$ is the CKM matrix element~\cite{pdg2020}, and
$m_\ell$ $\left[m_{D_{(s)}^+}\right]$ is the lepton $\left[D_{(s)}^+~\mathrm{meson}\right]$ mass.  Using the measured branching fractions of the leptonic $D_{(s)}^+$ decays,
the product $f_{D_{(s)}^+}|V_{cs(d)}|$ can be determined.
By taking the $f_{D_{(s)}^+}$ calculated by LQCD with precision of 0.2\%~\cite{prd98_074512,prd91_054507}, one can precisely determine the Cabibbo-Kobayashi-Maskawa (CKM) matrix elements $|V_{cs}|$ and $|V_{cd}|$, which are essential inputs to test the CKM matrix unitarity.
Conversely, taking the $|V_{cs}|$ and $|V_{cd}|$ from the standard model global fit,
one can precisely measure the $D_{(s)}^+$ decay constants, which are crucial to calibrate LQCD for heavy quark studies.
Comparing the obtained branching fractions of $D_{(s)}^+ \to \tau^+\nu_\tau$ and $D_{(s)}^+ \to \mu^+\nu_\mu$ gives
an important comprehensive test of $\tau-\mu$ lepton-flavor universality.

In recent years, BESIII reported the most precise experimental studies of $D_{(s)}^+ \to \ell^+\nu_\ell$ by using
2.93, 0.48, and 6.32 fb$^{-1}$ of data taken at $\sqrt s=3.773$, 4.009, and 4.178-4.226 GeV~\cite{bes3_muv,Ablikim:2019rpl,bes3_Ds_muv,bes3_Ds_tauv1,bes3_Ds_tauv2,bes3_Ds_tauv3}.
However, the statistical uncertainty is still dominated in the studies of $D^+ \to \ell^+\nu_\ell$
and the statistical and systematic uncertainties are comparable in those of $D^+_s \to \ell^+\nu_\ell$.
The full BESIII data samples to be collected in the coming years offer opportunities to further improve the precision of these important constants.
The relative uncertainties on
$f_{D^+}$ and $f_{D^+_s}$ can be reduced from 2.6\% and 1.2\% to approximately 1.1\% and 0.9\%, respectively,
where the statistical uncertainty will be dominated in the former one
while the systematic uncertainty is expected to be dominated in the latter one.
Similarly, the relative uncertainties in the measurements of $|V_{cs}|$ and $|V_{cd}|$ with leptonic decays
are expected to be reduced from 2.6\% and 1.2\% to approximately 1.1\% and 0.9\%, respectively.
Those of lepton-flavor universality test in $D^+ \to \ell^+\nu_\ell$ and $D_{s}^+ \to \ell^+\nu_\ell$ decays
are expected to be reduced from to 24.0\% and 4.0\% to about 10.0\% and 3.0\%, respectively.

\subsection{Semileptonic decays}
In the standard model, the weak and strong effects in semi-leptonic $D_{(s)}^{0(+)}$ decays
can be well separated~\cite{Ivanov:2019nqd}.
Among them, the simplest case is the semi-leptonic $D_{(s)}^{0(+)}$ decays into a pseudoscalar meson, e.g., $D^{0(+)}\to\bar{K}(\pi)\ell^+\nu_\ell$, for which the differential decay rate can be simply written as
\begin{equation}
\frac{d\Gamma}{dq^2} =\frac{G_F^2}{24\pi^3}|V_{cs(d)}|^2
p_{K(\pi)}^3 |f_{+}^{K(\pi)}(q^2)|^2,
\end{equation}
where $G_F$ is the Fermi coupling constant, and
$p_{K(\pi)}$ is the kaon (pion) momentum in the $D$ rest frame,
$f_{+}^{K(\pi)}(q^2)$ is the form factor of the hadronic weak
current depending on the square of the transferred four-momentum
$q=p_D-p_{K(\pi)}$.
From analyses of the dynamics in these decays,
one can obtain the product $f_{+}^{K(\pi)}(0)|V_{cs(d)}|$.
By taking the $f_{+}^{K(\pi)}(0)$ calculated in LQCD or $|V_{cd(s)}|$ from a
global fit assuming unitarity in the standard model,
the value of either $|V_{cs(d)}|$ or $f_{+}^{K(\pi)}(0)$ can be obtained.
The measurements with semi-leptonic $D$ decays offer the sub-leading contributions
of the world averages of $|V_{cs}|$ and $|V_{cd}|$, which are also important inputs to test the CKM matrix unitarity.
The obtained form factors $f_{+}^{K(\pi)}(0)$ are crucial to calibrate LQCD for heavy quark studies.
Comparing the branching fractions of the semi-muonic and semi-electronic $D$ decays
offers a comprehensive test of $\mu-e$ lepton-flavor universality.

Over the years, BESIII reported experimental studies of the semi-leptonic $D_{(s)}^{0(+)}$ decays into $P$, $V$, $S$, and $V$,
where $P$ denotes pseudoscalar mesons of
$K$~\cite{bes3_D0_kpiev,bes3_Dp_k0pi0ev,bes3_Dp_KLev,bes3_D_kev,bes3_Dp_KSev_2pi0,bes3_D0_kmuv,bes3_Dp_kmuv,bes3_Dst_Kev},
$\pi$~\cite{bes3_D0_kpiev,bes3_Dp_k0pi0ev,bes3_D_pimuv},
$\eta$~\cite{bes3_Dp_etaev,bes3_Dp_etamuv,bes3_Dst_etaenu,bes3_Ds_etaev_4009,bes3_Ds_etamuv_4009}, $\eta^\prime$~\cite{bes3_Dp_etaev,bes3_Dst_etaenu,bes3_Ds_etaev_4009,bes3_Ds_etamuv_4009};
$V$ denotes vector mesons of $K^*$~\cite{bes3_Dp_kpiev,bes3_D0_kspiev,bes3_Dst_Kev},
$\rho$~\cite{bes3_D0_pipiev,bes3_D0_rhomuv}, $\omega$~\cite{bes3_Dp_omegaev,bes3_Dp_omegamuv}, and $\phi$~\cite{bes3_Dp_omegaev,bes3_Ds_etamuv_4009};
$S$ denotes scalar mesons of $f_0$~\cite{bes3_D0_pipiev,bes3_Ds_f0enu} and $a_0$~\cite{bes3_D_a0ev,bes3_Ds_a0enu}; and
$A$ denotes axial vector mesons of $K_1$~\cite{bes3_Dp_K1enu,bes3_D0_K1enu} and $b_1$~\cite{bes3_D_b1enu}.
These measurements were carried out by using 2.93, 0.48, and 6.32 fb$^{-1}$ of data taken at $\sqrt s=3.773$, 4.009, and 4.178-4.226 GeV.

Except for the $D^{0(+)}\to K$ and $D^{0(+)}\to K^*$ form factors, the precision of all other measurements of
the $D_{(s)}^{0(+)}\to P$ and $D_{(s)}^{0(+)}\to V$ form factors are restricted due to limited data sets.
With full BESIII data samples, all the form-factor measurements which are currently statistically limited
will be statistically improved by a factor of up to 2.6
and 1.4 for semi-leptonic $D^{0(+)}$ snd $D^+_s$ decays, respectively.
In addition, studies of the semi-muonic charmed meson decays further improve the form factor knowledge.
We hope to extract the $D\to S$ and $D\to A$ form factors for the first time.

The best precision of the $c\to s$ and $c\to d$ semi-leptonic $D^{0(+)}$ decay form factors
will be from the studies of $D^{0(+)}\to \bar K\ell^+\nu_\ell$ and $D^{0(+)}\to \pi\ell^+\nu_\ell$.
Combining analysis of semi-electronic and semi-muonic $D^0$ and $D^+$ decays will give more precise results.
The experimental uncertainties are expected to be reduced from 0.6\% and 1.5\% to the level of 0.4\% and 0.7\%, respectively.
While the best precision of $c\to s$ and $c\to d$ semi-leptonic $D^+_s$ decay form factors
will be from the studies of $D^+_s\to \eta \ell^+\nu_\ell$ and $D^+_s\to K^0\ell^+\nu_\ell$.
The experimental uncertainties are expected to be reduced from 4.0\% and 10.0\% to the level of 2.0\% and 5.0\%, respectively.
In addition, the dynamics studies of the other semi-leptonic $D\to P$ decays will offer comprehensive measurements of
$|V_{cs}|$ and $|V_{cd}|$.

Currently, the precisions of the measurements of $|V_{cs}|$ and $|V_{cd}|$ with semi-leptonic $D^{0(+)}$ decays
are limited by the theoretical uncertainties of $f^K_+(0)$ and $f^\pi_+(0)$ in LQCD, which
are 2.4\% and 4.4\%, respectively~\cite{lqcd_fk,lqcd_fpi,lqcd_ETM,lqcd_MILC}.
In the coming decades, if the uncertainties of $f^K_+(0)$ and $f^\pi_+(0)$ calculated in LQCD can be reduced to be better than the level of 1.0\%, the experimental uncertainties of the measurements of $|V_{cs}|$ and $|V_{cd}|$ with semi-leptonic charmed meson decays
are hopefully to reach about 1.0\% level.

For semi-leptonic $D^{0(+)}_{(s)}$ decays, the best test of $\mu-e$ lepton-flavor universality is expected to be
from $D\to \bar K\ell^+\nu_\ell$ decays, where the test precision can be reduced from 1.3\% to the level of 0.8\% in the near future.
At present, it is not conclusive about whether the $\mu-e$ lepton-flavor universality always holds in semi-leptonic $D^{0(+)}_{(s)}$ decays,
because there are still many un-observed semi-muonic decays, e.g.,
$D^+\to \eta^\prime \mu^+\nu_\mu$, $D^{0(+)}\to a_0(980) \mu^+\nu_\mu$,
$D^{0(+)}\to K_1(1270) \mu^+\nu_\mu$, $D^+\to f_0(500) \mu^+\nu_\mu$, $D^+_s\to K^0\mu^+\nu_\mu$, $D^+_s\to K^{*0}\mu^+\nu_\mu$,
$D^+_s\to f_0(980)\mu^+\nu_\mu$, and $D^+_s\to \eta^\prime \mu^+\nu_\mu$.
Larger data samples will offer opportunities to search for these decays and
thereby clarify if there is possible violation of $\mu-e$ lepton-flavor universality in the charm sector.

Moreover, the studies on the intermediate resonances in hadronic final states, e.g., $K_1(1270)$ and $a_0(980)$, in
the semi-leptonic $D^{0(+)}_{(s)}$ decays provide a clean environment to explore meson spectroscopy, as no other particles interfere.
This corresponds to a much simpler treatment than those studies in charmonium decays or hadronic $D^{0(+)}_{(s)}$ decays.

\subsection{Quantum-correlated $D$ decays}
Another way to test precisely CKM unitarity, as well as $CP$ violation within the SM, is 
to improve the measurements of the CKM Unitarity Triangle angles of $\alpha$,
$\beta$ and $\gamma$, also known as $\phi_2$, $\phi_1$ and $\phi_3$, in $B$ decays.
Among the three CKM angles, $\gamma$ is the only one that can be determined using tree-level decays only. 
The independence from loop diagrams means the measurements have negligible theoretically uncertainty \cite{BROD}.
The precision measurement of $\gamma$ is one of the top priorities for both the LHCb, LHCb upgrade and Belle II experiments.

Currently the most precise method to measure $\gamma$ is
based on the interference between $B^{+}\to\bar{D}^{0}K^{+}$ 
and $B^{+}\to D^{0}K^{+}$ decays, where the $D^0$ or $\bar{D}^0$ decays to $K^0_{\rm S}\pi^+\pi^-$ ~\cite{GGSZ}. 
In the future, the statistical uncertainties of these measurements
will be greatly reduced by using the huge $B$ samples produced by Belle II and LHCb and by
extending these measurements to other similar $B$ modes. 
However, with increased statistical precision, the knowledge of strong-phase measurements of the $D$ decays will systematically limit the uncertainty. Consequently, the improved knowledge of the strong-phase  related parameters in $D$ decays is essential to making measurements of $\gamma$ to degree level precision. The current BESIII strong-phase measurement in $D^0\to K_{\rm S}^0\pi^+\pi^-$ \cite{BESIIIcisi}, based on $2.93$~fb$^{-1}$ of data, induces a systematic uncertainty of $1^{\circ}$-$1.4^{\circ}$ on $\gamma$ \cite{lhcb:gamma,belle:gamma}. The measurements of the strong-phase differences are not expected to be statistical limited so this is expected to reduce to $0.4^{\circ}$ when measured in a 20~fb$^{-1}$ data set. This systematic uncertainty will not be limiting the precision on $\gamma$ for the foreseeable future.  

The current best measurements of the charm-mixing parameter $x$ and constraints on indirect {\it CP} violation in charm come from measurements of $D^0\to K^0_{\rm S}\pi^+\pi^-$ \cite{binflip} relies on strong-phase measurements. The largest systematic uncertainty comes from the related uncertainties. So these measurements will also benefit from the improved precision provided by a 20~fb$^{-1}$ data set. 

Many other strong-phase measurements are used for $\gamma$ and charm-mixing measurements. These measurements are unique to BESIII; more details can be found in Ref. \cite{whitepaper}.

\subsection{Hadronic $D$ decays}

\label{sec:precision_d}

%\subsection{Precision measurements of absolute branching fractions}

Some experiments, for example LHCb, have the ability to measure a large number 
of charm and beauty hadron relative branching-fraction ratios due to the high yields 
given by the large charm and beauty production cross section.
The conversion from the branching-fraction ratio to the absolute branching fraction
incurs the uncertainty of the branching fraction of 
the reference mode, such as, 
$D^0\to K^-\pi^+$, $D^0\to K^-\pi^+\pi^+\pi^-$, $D^+\to K^-\pi^+\pi^+$,
$D^+_s\to K^-K^+\pi^+$, and $\Lambda_c^+\to pK^-\pi^+$.
Improved measurements of these absolute branching fractions at BESIII
will be highly beneficial to some key measurements at LHCb,
since it is expected that the uncertainty of the reference mode 
will become the dominant uncertainty in several measurements.
With 20~fb$^{-1}$ data taken around $\sqrt s=$ 3.773 and 4.18 GeV at BESIII,
these decays are expected to be measured with an uncertainty of about 1\%.

At present, the sum of the branching fractions for the known
exclusive decays of $D^0$, $D^+$ and $D^+_s$ are more than 80\%~\cite{pdg2020}.
However, there is still significant room to explore more hadronic decays  
to increase the known branching fractions for $D^0$, $D^+$ and $D^+_s$.
A 20 fb$^{-1}$ dataset will allow the determination of the absolute branching fractions of those 
missing decays $K\pi\pi\pi$, $KK\pi\pi$, and $KK\pi\pi\pi$ and exploring the sub-structures in these decays using amplitude analyses is also interesting.
In addition, precise measurements of the branching fractions for 
$D^0$, $D^+_s$ and $D^+$ inclusive decays to three charged pions and
other neutral particles, and exclusive decays to final states with neutral kaons and pions
(e.g. $D_s^+\to \eta^\prime\pi^+\pi^0$, $D^+\to \bar K^0\pi^+\pi^+\pi^-\pi^0$ 
and decay modes contributing to $D^{0(+)}\to\eta X$) are also highly desirable to better understand backgrounds in several measurements, particularly $B\to D^{*}\tau^+\nu_\tau$.

Studies of such multi-body decays benefit from amplitude analyses to understand the intermediate resonances. Even though it is possible to accumulate large samples of  singly tagged $D$ mesons, they have very high backgrounds making them unsuitable to perform amplitude analyses. In contrast to this, the doubly tagged $D\bar D$ mesons can  provide clean $D$ samples with low backgrounds, but suffering  from limited statistics.

\subsection{Charmed baryons}

The lightest charmed baryon, $\Lambda_c^+$, which was observed in 1979,
is the cornerstone of the charmed baryon spectra.
The improved knowledge of $\Lambda_c^+$ decays,
especially for the normalization mode $\Lambda_c^+\to pK^-\pi^+$,
is key for the studies of the charmed baryon family.
Moreover, the $\Lambda_c^+$ decays can also open a window upon a 
deeper understanding of strong and weak interactions in the charm sector.
In addition, these will provide important inputs for the studies of
beauty baryons that decay into final states involving $\Lambda_c^+$. 

Compared to the significant progress in the study of charmed 
mesons ($D^0$, $D^+$ and $D^+_s$), the advancements in 
the knowledge of the charmed baryons are relatively slow in the 
past 40 years.
Before 2014, almost all the decays of $\Lambda_c^+$ were measured relative 
to the normalization mode $\Lambda_c^+\to pK^-\pi^+$, which branching 
fraction suffered a large uncertainty of 25\%. 
Moreover, no data sample taken around the 
$\Lambda_c^+\bar \Lambda_c^-$ pair production threshold had been
used to study the $\Lambda_c^+$ decays. 

We have already collected 4.4~fb$^{-1}$ of data above $\Lambda_c\bar{\Lambda_c}$ threshold, which will provide the most precise values of many branching fractions and polarization parameters \cite{whitepaper}. Future running with the upgraded BEPC-II will allow large samples of $\Sigma_c$ and $\Xi_c$ pairs to be collected, which will allow absolute branching fractions of many charm baryon decays to be determined for the first time \cite{whitepaper}.

\section{R Values, QCD, and tau Physics}
BESIII is able to measure the $e^+e^-$ annihilation cross section
from threshold to the maximum BEPCII energy 
using a combination of initial state radiation and fine energy scans.
The data allow
fundamental physics measurements of the hadronic vacuum
polarization (HVP), crucial input to hadronic corrections
to the muon $g-2$ and to the fine structure constant $\alpha_{\rm
EM}(s)$. Beyond HVP, the next important contribution to the
uncertainty of $(g-2)_\mu$ is given by the HLbL contribution. The
leading contribution to the HLbL diagram is given by the coupling
of photons to the pseudoscalar mesons $\pi^0$, $\eta$,
$\eta^\prime$ as well as channels like $\pi\pi$ and $\pi\eta$ in
the region of small momentum transfers, exactly where
BESIII can provide precision results.
The same data provides access to time-like baryon form factors, where a number of surprising features have been discovered.  Among them, the sharp rise of
baryon-antibaryon cross sections close to threshold appears very intriguing.
BESIII has initiated a program to systematically study this effect
for all accessible baryons.
In addition, the $\tau$ mass can be
measured at BESIII with a precision reaching 0.1~MeV, providing a
stringent test of lepton universality in the SM.  This measurement uses the beam energy measurement system (BEMS) based on the
back-scattered Compton photon energy spectrum, allowing beam energy measurements reaching a precision of a few 10~keV.

% RM EDITED: 4/19
\subsection{Hadronic cross sections and transition form factors}
The recent measurement of $(g_\mu-2)/2 = a_\mu$ at Fermilab~\cite{Muong-2:2021ojo} 
has confirmed the long-standing discrepancy between the Standard Model (SM) prediction~\cite{Aoyama:2020ynm} 
of the muon $g-2$ and the direct measurement at the level of $4.2\sigma$. However, the SM prediction is limited by uncertainties due to the contributions of the strong interaction, namely the hadronic vacuum polarization (HVP) and the hadronic light-by-light scattering (HLbL). A comparable limitation due to HVP effects exists for predictions of the running of the fine-structure constant at the $Z$-pole, $\alpha_{\rm em}(M_Z)$, %~\cite{},
which is essential for electro-weak precision physics. Data-driven approaches connect the hadronic contributions to experimental data so that the predictions can be systematically improved by more precise measurements. The HVP contributions are linked to hadronic cross sections in $e^+e^-$ collisions, while the HLbL contribution is linked to the transition form factors of mesons.

% RM EDITED: 4/19
\subsubsection{Exclusive cross section measurements using initial state radiation}
The dispersive integral formalism used to determine the HVP contribution to $a_\mu$ relies heavily on the hadronic $e^+e^-$ cross sections at CM energies $\sqrt{s}\leq2$\,GeV. At BESIII, these energies are only accessible by exploiting the initial state radiation (ISR) method, where events with an additional hard photon emitted from the initial state are selected. The radiative photons can be reconstructed at large angles in the detector, or at small angles using energy and momentum conservation, allowing access to the full differential cross section of the ISR photon emission, albeit with finite acceptance only for multi-meson systems with masses above 1\,GeV. 
With an initial data set of $2.83\,\textrm{fb}^{-1}$ at $\sqrt{s}=3.773$\,GeV,
this technique already produces results competitive with the B-factories for hadronic masses above approximately 1.3\,GeV.

In a first measurement by BESIII, the largest hadronic cross section, for $e^+e^-\to\pi^+\pi^-$, was measured in the mass region from 600 to 900\,MeV by reconstructing the ISR photon at large angles only~\cite{BESIII:2015equ}. 
The most challenging aspect of the measurement was the rejection of background contributions from muon production, which was achieved using machine learning techniques. The resulting cross section was obtained with an accuracy of 1\%, 
where the largest uncertainty is due to the integrated luminosity and the theoretical uncertainty on the radiator function.
%Though the result favors one of the two leading measurements, its accuracy does not allow to settle the discrepancy between these results. 
With 20\,fb$^{-1}$ of data at $\sqrt{s}=3.773$\,GeV expected soon, a new measurement of the $\pi^+\pi^-$ cross section will use the improved statistical accuracy to implement an alternative normalization scheme relative to the muon yield. With this approach, the largest uncertainties will cancel, bringing the expected final uncertainty down to 0.5\%.

Additionally, the multi-meson cross sections for $e^+e^-\to\pi^+\pi^-\pi^0$~\cite{BESIII:2019gjz} as well as $e^+e^-\to\pi^+\pi^-\pi^0\pi^0$~\cite{Redmer:2018nxj} have been measured using the same analysis strategy. In both cases, the ISR photon was reconstructed at large as well as at small angles, and the results were averaged.  Uncertainties of approximately 3\% were achieved, which is competitive with the existing measurements from the BaBar collaboration above 1.3\,GeV. These cross sections can be used 
to study resonances in the final state as well as in the intermediate states. Further improvements are expected with additional data at $\sqrt{s}=3.773$\,GeV.

% RM EDITED (loosely): 4/19
\subsubsection{Inclusive R measurement}
The R ratio, defined as the lowest-order cross section for inclusive hadron production, $e^{+}e^{-}\to hadrons$, normalized by the lowest-order cross section for the QED process $e^{+}e^{-}\to\mu^{+}\mu^{-}$, is a central quantity in particle physics. Precision measurements of the R ratio below 5~GeV contribute to the MC prediction of the muon anomalous magnetic moment. The R ratio also contributes in the determination of the QED running coupling constant evaluated at the Z pole.  In a first measurement at BESIII~\cite{BESIII:2021wib}, 14 data points with CM energies from 2.2324 to 3.6710~GeV are used for the inclusive R value measurement. Taking advantage of the large data samples, the statistical uncertainty of the measured R ratio is less than 0.6\%. Two different simulation models are used in the analysis. On the one hand, the established LUARLW generator, based on a purely theoretical description of the hadronization process, and on the other hand, a newly constructed hybrid generator, making use of as much experimental information as possible\cite{Ping:2016pms}, are used, which reproduce the inclusive hadronic event production from electron-positron annihilation very well. The two simulation models give consistent detection efficiencies and initial-state-radiation corrections, with a maximum total difference of less than 2.3\%. An accuracy of better than 2.6\% below 3.1 GeV and 3.0\% above is achieved in the R ratios, much improved from previous results at the level of 3-6\%.  

The complete data set for the R value measurement at BESIII consists in total of 130 energy points with an integrated luminosity of about 1300\,pb$^{-1}$, corresponding to more than $10^5$ hadronic events at each of the points between 2 and 4.6\,GeV. Thus, the final result is expected to be dominated by a systematic uncertainty of less than 3\%.

\subsubsection{Meson transition form factors}
Transition form factors (TFF) of mesons $M$ describe the effects of the strong interaction on the $\gamma^*\gamma^*M$ vertex. They are represented by functions $F_{M\gamma^*\gamma^*}(q_1^2,q_2^2)$ of the virtualities $q_1^2$ and $q_2^2$ of the photons. The TFF provide information on the structure of the mesons and are essential inputs to the data-driven evaluations of the hadronic light-by-light contribution to $a_\mu$. At BESIII the TFF can be studied in the region of time-like virtualities through meson Dalitz decays and the radiative meson production in $e^+e^-$ annihilations. In the former the virtualities are limited by the meson mass and in the latter the virtuality is fixed to the CM energy. 
Space-like virtualities are studied in two-photon fusion reactions, which in principle allow to map the TFF over a wide region of virtualities by measuring the momentum transfer of the scattered electrons. Due to the rapid drop of the cross section with $Q_i^2=-q_i^2$ the investigations at BESIII are currently done as single-tagged measurements, where the TFF is only studies depending on one of the virtualities. A first measurement of the $\pi^0$ TFF based on 2.83\,fb$^{-1}$ of data at $\sqrt{s}=3.773$\,GeV covers virtualities from 0.3\,GeV$^2$ to 3.1\,Gev$^2$, with unprecedented accuracy for $Q^2\leq1.5$\,GeV$^2$. The results confirm the recent calculations in disperion theory %~\cite{} 
and on the lattice. %~\cite{}.
The range of virtualities covered by BESIII is complementary to the results of the B-factories. %~\cite{}. 
It is limited to smaller values by the acceptance of the detector for the decay products and towards larger values by statistics.

Analogous studies are performed for $\eta$ and $\eta^\prime$ mesons, and also for multi-meson systems. The production of charged and neutral two-pion systems in two-photon fusion gives access to pion masses from threshold to 2\,GeV and virtualities from 0.2\,Gev$^2$ to 3\,Gev$^2$ at a full coverage of the pion helicity angle. 
The results will be complementary to all previous measurements, which have mostly been performed with quasi-real photons, i.e. $Q^2=0$\,GeV$^2$ only, apart from the recent measurement of $\pi^0\pi^0$ production by Belle, %~\cite{}, 
providing data only for $Q^2>4\,\textrm{GeV}^2$.

The production of higher meson multiplicities in two-photon fusion allows access to scalar, tensor and axial resonances. The single-tagged strategy allows for the production of axial mesons due to the presence of a highly virtual photon. A first measurement of the $f_1(1285)$ will be performed using the $\pi^+\pi^-\eta$ final state for reconstruction. Virtualities from 0.2\,Gev$^2$ to 3\,Gev$^2$ can be covered and in contrast to previous measurements directly measured from the scattering products. Additionally, the dominating intermediate state $a_0(980)\pi$ allows to separate the different helicity states of the $f_1(1285)$ TFF base on the angular distributions of the decay products.

With the upcoming data set of 20\,fb$^{-1}$ at $\sqrt{s}=3.773$\,GeV all two-photon fusion analysis will benefit from higher statistics, which will allow access to higher virtualities. A direct comparison to results from B-factories can become possible. Based on existing data sets of the R value, measurements below $\sqrt{s}=3$\,GeV virtualities down to $Q^2=0.1\,\textrm{GeV}^2$ are accessible. The combination of the large present and future data sets at BESIII will also allow for the measurements of the full TFF of different mesons depending on both virtualities.

\subsection{Search for light-flavor vector mesons}
There are rich vector resonances around 2.0~GeV observed experimentally, interpreted as excited $\rho$, $\omega$ and $\phi$ resonances,
but their properties are controversial. 
The $\phi(2170)$ has drawn  a lot attention theoretically since it could be an exotic state candidate. 
%Due to its controversial property, more experimental  researches are necessary.   
At BESIII, the line-shapes of $e^{+}e^{-}\to\phi\eta'$~\cite{BESIII:2020gnc} and $\phi\eta$~\cite{BESIII:2021bjn} are measured from 2.0 to 3.08 GeV.
A resonance structure are observed in both processes around 2.175~GeV, and it could be interpreted as a $\phi(2170)$   
due to rich $s\bar{s}$ component. 
%and resonance structures are observed around 2.175~GeV in both processes. The mass and width of the structures are obtained and they
%could be a possible $\phi(2170)$ candidate 
Similar resonance structures are observed in $e^{+}e^{-}\to K^{+}K^{-}$~\cite{BESIII:2018ldc} and $K_{S}K_{L}$~\cite{BESIII:2021yam} lineshapes.
However, the mass of the resonance from the two processes are  different from the average PDG value of $\phi(2170)$. 
Multiple line-shapes of intermediate state $e^{+}e^{-}\to K K_{1}(1400)$, $K K(1460)$, $K^{*}(892)K^{*}(892)$
and $K K_{1}(1270)$
are obtained by a partial-wave analysis of $e^{+}e^{-}\to K^{+}K^{-}\pi^{0}\pi^{0}$~\cite{BESIII:2020vtu}.
The structures observed in these intermediated processes are fitted simultaneously and the partial width of each process is determined.
Similarly,  a partial-wave analysis is performed for the process $e^{+}e^{-}\to K^{+}K^{-}\pi^{0}$~\cite{BESIII:2022wxz} where the intermiediated processes
$e^{+}e^{-}\to\phi\pi^{0}$, $e^{+}e^{-}\to K^{*+}(892)K^{-}$, and $K_{2}^{*+}(1430)K^{-}$ are measured. 
These results provide essential input to understand the nature of $\phi(2170)$.
The line-shapes of $e^{+}e^{-}\to\omega\pi^{0}$~\cite{BESIII:2020xmw} and $\eta'\pi^{+}\pi^{-}$~\cite{BESIII:2020kpr} are studied and the structures observed could be
excited $\rho$ resonances, such as $\rho(2000)$ or $\rho(2150)$.
The line-shape of $e^{+}e^{-}\to\omega\pi^{0}\pi^{0}$~\cite{BESIII:2021uni} is measured and a resonant structure around 2.20~GeV is observed which could be an excited $\omega$ resonances.

\subsection{Time-like baryon form factors}
%Proton and neutron, refereed as nucleons, are the fundamental building blocks of the universe. 
%Though been discovered over one century, the structure of the nucleon is still not fully understood. 
The simplest structure observable of nucleons are the electromagnetic form factors~(EMFFs)
that relate to their charge and magnetization distribution, and provide crucial testing ground for QCD-related models.
%EMFFs are the functions of the four momentum transfer squared~($q^{2}$) carried by the exchanged virtual photon. 
There are two FFs for spin 1/2 baryons, namely electric $G_{E}$ and magnetic $G_{M}$ FFs. 
The EMFFs have been studied in space-like $(q^{2}<0)$ by electron-nucleon scattering experiment extensively~\cite{Punjabi:2015bba}
with uncertainties of $\mu_{p}|G_{E}/G_{M}|$ to be 1.7\% for proton~\cite{Puckett:2011xg}. 
However, precision of EMFFs in time-like $(q^{2}>0)$ by electron-positron collider experiments was not compatible.
At BESIII, both energy scan method~\cite{BESIII:2015axk,BESIII:2019hdp} and ISR approach~\cite{BESIII:2019tgo,BESIII:2021rqk} have been applied for the study of proton EMFFs.
The $|G_{E}/G_{M}|$ of proton is determined over a large $q^{2}$ from threshold to 9.5~GeV$^{2}$ with the best precision achieving 3.7\%.
Model-independent line-shapes of $|G_{E}|$ and $|G_{M}|$ are determined for the first time. 
The neutron EMFFs have been investigated at BESIII by energy scan method via $e^{+}e^{-}\to n\bar{n}$ process~\cite{BESIII:2021tbq}.
The cross section of $e^{+}e^{-}\to n\bar{n}$ is found to be smaller than that of $e^{+}e^{-}\to p\bar{p}$ which
indicates the photon-proton interaction is stronger than the corresponding photon-neutron interaction, as expected by 
most theoretical predictions~\cite{Chernyak:1983ej,Ellis:2001xc}, in contrast to the results obtained by the FENICE experiment~\cite{Antonelli:1998fv}.
Due to limited statistics, the electric and magnetic FFs of neutron cannot be separated and the effective FFs $|G_{\rm eff}|$ is obtained.
The effective FFs of neutron show a periodic behavior, similar to earlier observations of proton FFs reported by BaBar~\cite{BaBar:2013ves}. 
The results imply that there are some not yet understood intrinsic dynamics that are responsible for almost orthogonal oscillations. 
More high-precision data are needed to establish this elusive feature. Besides, the disentanglement of neutron's EMFFs can be achieved with more statistics which is of great importance for a comprehensive understand of nucleon structure.

Hyperon EMFFs are hard to probe in space-like since they are unstable as beams or targets.
Instead, their structures can be studied in the time-like region via the electron-positron annihilation. 
The energy region of BESIII covers the production threshold of all SU(3) octet hyperons and several charmed baryons.    
At BESIII, the Born cross sections of electron-positron annihilation to various baryon pairs
 are measured from threshold, including $\Lambda\bar{\Lambda}$~\cite{BESIII:2017hyw}, 
$\Sigma\bar{\Sigma}$~\cite{BESIII:2020uqk,BESIII:2021rkn}, $\Xi\bar{\Xi}$~\cite{BESIII:2020ktn,BESIII:2021aer} and $\Lambda_{c}\bar{\Lambda}_{c}^{+}$~\cite{BESIII:2017kqg}.
The $|G_{E}/G_{M}|$ of $\Lambda$, $\Sigma^{+}$ and $\Lambda_{c}$ are obtained with angular analysis while effective FFs are extracted for other baryons. 
Under one-photon exchange approximation~\cite{Cabibbo:1961sz}, the production cross sections are expected to be zero at threshold 
for neutron baryon pairs due to vanishing phase space and non-zero for charged baryon pairs due to the 
the Sommerfeld resummation factor~\cite{Arbuzov:2012nm}, and then increases with phase space above threshold.
However, anomalous threshold effects are observed where a sudden rise of 
$e^{+}e^{-}\to\Lambda\bar{\Lambda}$~\cite{BESIII:2017hyw} cross section and a flat platform of $e^{+}e^{-}\to\Lambda_{c}\bar{\Lambda}_{c}^{+}$~\cite{BESIII:2017kqg}
 are observed. For other baryon pairs, no such threshold effects observed. 
More precise data or finer scan are necessary for a deeper insight of these results. 
The hyperon EMFFs and the cross section line shapes can also be studied with improved precision
via ISR approaches with a  20~fb$^{-1}$  data set collected at $\sqrt{s}=3.773$~GeV.
There is a discrepancy of the $e^{+}e^{-}\to\Lambda_{c}\bar{\Lambda}_{c}^{+}$ cross section near threshold
 between Belle~\cite{Belle:2007umv} and BESIII. Currently, BESIII has taken data from $\sqrt{s}=4.6$ to 4.95~GeV for an extension study of
 $e^{+}e^{-}\to\Lambda_{c}\bar{\Lambda}_{c}^{+}$ to further review the resonance near 4.63~GeV observed by Belle. 
 
The EMFFs in time-like are complex that the relative phase between $G_{E}$ and $G_{M}$ will lead to the transverse polarization
of final baryons. Due to the weak decays of hyperons, the polarization can be measured by analyzing the angular of daughter baryon
from hyperon decay, so deoes the relative phase $\Delta\Phi$ between $G_{E}$ and $G_{M}$.
At BESIII, the relative phase of $\Lambda$ is determined at $\sqrt{s}=2.396$~GeV with a joint angular distribution analysis,
to be $\Delta\Phi=37^{\circ}\pm12^{\circ}\pm6^{\circ}$~\cite{BESIII:2019nep}. Combine the obtained $|G_{E}/G_{M}|$ at the same CM~energy, 
the complete EMFFs is determined for the first time. 
Similarly, relative phase of $\Lambda_{c}$ is determined at $\sqrt{s}=4.60$~GeV~\cite{BESIII:2019odb}.
The currently available data set from $\sqrt{s}=4.6$ to 4.95~GeV will help complete determinations of $\Lambda_{c}$ EMFFs in a wide $q^{2}$ range.
As the energy dependence of relative phase is essential for distinguishing various theoretical predictions, 
a complete determination of EMFFs for SU(3) octet hyperons are necessary in future. 

\subsection{Fragmentation functions}
The fragmentation function describes the probability of a hadron to be found in the debris of a quark, carrying a fraction of the quark. 
Precise knowledge of fragmentation functions could help to understand the internal structure of nucleon studied in the semi-inclusive deep inelastic scattering~(SIDIS) experiments at ${\it e.g.}$ the Electron-Ion Collider. 
Fragmentation functions are assumed to be universal, thus a large amount of data on inclusive hadron
production from $e^{+}e^{-}$ collisions has been collected in a wide energy range $10\leq\sqrt{s}\leq 200$~GeV~\cite{Arleo:2008dn}.
However, results from GeV energy scale are rare with large uncertainty. 
BESIII offers a unique opportunity to extract the unpolarized fragmentation functions in the
relatively low energy region and to study the QCD dynamics in this particular region.
At BESIII, with the data collected
at the continuum energy region, unpolarized fragmentation functions can studied from inclusive hadron production $e^{+}e^{-}\to h+X$, where $h$ denotes $\pi^{0}$, $\eta$, $K_{S}$ or charged hadrons.

The Collins fragmentation function,  which describes the spin-dependent effects in fragmentation processes, is an important input for the extracting of transversity distribution inside the nucleon of SIDIS experiment~\cite{Collins:1992kk}.
The energy scale of BESIII is close to that of existing SIDIS experimental data ($2\sim20$~GeV$^2$) and results at BESIII is crucial for exploring the $Q^{2}$ evolution of Collins fragmentation function.
The Collins asymmetries for $\pi$ pairs have been obtained by BESIII using the data collected at $\sqrt{s}=3.65$~GeV~\cite{BESIII:2015fyw}.
A nonzero asymmetry is observed which increases with increasing $\pi$ momentum. 
The BESIII asymmetry in the large $\pi$ momentum interval is about
two-three times larger than measured at the B factories.
Additional results are needed to confirm this observation. 
The Collins effect for strange quarks could be studied in $e^{+}e^{-}\to \pi K+X$ and $e^{+}e^{-}\to KK+X$. It is also interesting to study the Collins effect in neutral hadrons like $e^{+}e^{-}\to P P^{'}+X$ with $P/P^{'}=\pi^{0}/\eta$.

\subsection{Relative phase in vector charmonium decays}
Studying the relative phase between strong and EM amplitudes of narrow charmonium decay provides
us new directions to explore decay dynamics of quarkonium system.  
At BESIII,  studies of phase difference have been performed with $J/\psi$, $\psi(2S)$ scan data. 
A relative phase of $|\Delta\Phi|\sim90^{\circ}$ of $J/\psi\to\eta\pi^{+}\pi^{-}$ was observed~\cite{BESIII:2018wid}, 
and there are a series of processes under investigation, {\it e.g.} $\psi\to KK$ and various baryon pairs. 
In $\psi$ decays into baryon pairs, the “magnetic”
and “electronic” amplitudes might have different phases.

An intriguing situation has been found in $J/\psi\to\pi\pi$.
Due to the $G$ parity violation, a purely strong amplitude
should be suppressed. However, the EM amplitude is not
sufficient to explain the partial with of $J/\psi\to\pi\pi$. As a
remedy, an amplitude was suggested with two gluons and
one virtual photon as an intermediate state~\cite{BaldiniFerroli:2016mbs}.
In order to test this hypothesis, it is most interesting to determine the relative phase between the strong and EM decay
amplitudes in the decay. The currently available data at BESIII, however, do not allow such investigations.
Therefore, more data are needed to settle
the issue of the phase difference between the strong and
EM decay amplitudes of vector charmonium states.
 
% RM EDITED: 4/15
\subsection{Physics of the $\tau$ lepton}
Since the discovery of the $\tau$ lepton in 1975 at the SPEAR $e^{+}e^{-}$ storage ring~\cite{Perl:1975bf}, the study of the $\tau$ lepton has been an extensive experimental subject.  The properties of the $\tau$ lepton, including its mass, lifetime, and decays, have been tested with increasing rigor~\cite{Asner:2009zza}. As a member of the third fermion generation, it decays to fermions in the first and second generations. The pure leptonic or semi-hadronic character of $\tau$ decays provides a clean laboratory to test the structure of the weak currents and the universality of their couplings to the gauge bosons. Moreover, the $\tau$ is the only lepton massive enough to decay into hadrons, and its semi-hadronic decays are an ideal tool for studying strong interactions. The great advantage of BEPCII lies in running near the threshold for $\tau$ pair production, which provides an excellent opportunity for $\tau$ lepton physics. Compared with other experiments, the background is simple and the systematic uncertainties can be easily controlled. The energy spread of the BEPCII beam is small, about 1-2 MeV. Moreover, the beam energy measurement system (BEMS) was built to determine the energy and energy spread accurately.

% RM EDITED: 4/15
\subsubsection{Precision measurement of the $\tau$ mass}
The $\tau$ lepton is one of three charged elementary leptons in nature, and its mass is an important parameter of the Standard Model. The $\tau$ mass can and should be provided by experiment precisely.  Compared to the other two leptons, the electron and the muon, the precision on the $\tau$ mass is almost four orders of magnitude worse. Precision $\tau$ mass measurements probe lepton universality, which is a basic ingredient in the Standard Model.

In order to minimize the statistical uncertainty on the $\tau$ mass measurement for a fixed integrated luminosity, a study was carried out using MC simulation to find the optimal data taking scheme, including how many points should be taken, the position of each point, and the allocation of the integrated luminosity among these points.  Since the beam energy uncertainty is the dominant source of systematic uncertainty for the $\tau$ mass measurement, a high-accuracy beam energy measurement system~(BEMS), located at the north crossing point of BEPCII, was designed, constructed, and finally commissioned at the end of 2010. By comparing a $\psi(2S)$ scan result with the PDG value of the $\psi(2S)$ mass, the relative accuracy of the BEMS was determined to be at the level of $2 \times 10^{-5}$~\cite{Abakumova:2011rp}.

Based on these preparations, the BESIII collaboration performed a fine mass scan experiment in the spring of 2018. In order to determine the beam energy, the $J/\psi$ and $\psi(2S)$ resonances were scanned at seven and nine energy points, respectively. The $\tau$ mass scan data were collected at five scan points near the pair production threshold with total luminosity of 137~pb$^{-1}$. The analysis is in progress. The uncertainty, including statistical and systematic error, will be less than 0.1 MeV.

% RM EDITED: 4/15
\subsubsection{Measurement of branching fraction of $\psi(2S) \to \tau \tau$}
The $\psi(2S)$ provides a unique opportunity to compare the three lepton generations by studying the leptonic decays $\psi(2S) \to e^+e^-$, $\mu^+\mu^-$, and $\tau^+\tau^-$. According to the Standard Model, the sequential lepton hypothesis leads to a relationship between the branching fractions of these decays, $B_{e^+e^-}$, $B_{\mu^+\mu^-}$, and $B_{\tau^+\tau^-}$, given by:
\begin{equation}
\frac{B_{e^+e^-}}{v_e (\frac{3}{2}-\frac{1}{2} v_e^2)}=
\frac{B_{\mu^+\mu^-}}{v_\mu (\frac{3}{2}-\frac{1}{2} v_\mu^2)}=
\frac{B_{\tau^+\tau^-}}{v_\tau (\frac{3}{2}-\frac{1}{2} v_\tau^2)},
\label{Eq.fracrelation}
\end{equation}
where $v_l=[1-(4m_l^2/M^2_{\psi(SS)})]^{1/2}$, $l=e,\mu,\tau$ is the velocity of the lepton. Substituting the lepton mass values yields:
\begin{equation}
B_{e^+e^-} \approx B_{\mu^+\mu^-} \approx \frac{B_{\tau^+\tau^-}}{0.3885}\equiv B_{ll}.
\label{Eq.fracthree}
\end{equation}
With the large $\psi(2S)$ data sample at BESIII, the relation between the three leptons can be tested by improving the accuracy of the branching ratio of $\psi(2S) \to \tau^+\tau^-$. As a byproduct, the total number of $\tau$ pairs can be used as an input into other analyses of $\tau$ physics.

% RM EDITED: 4/15
\subsubsection{Semi-hadronic decay of the $\tau$ lepton}
The $\tau$ lepton is the only lepton heavy enough to decay into hadrons. 
In fact, its partial decay width to final states including hadrons is about 65\%. Its semi-hadronic decays could help us understand the weak interactions and the mechanisms for hadronization in $\tau$ decays. 
A calculation of the semi-hadronic decay $\tau \to M_1 M_2 \nu_{\tau}$ using the basic weak interaction and angular momentum algebra has been performed~\cite{Dai:2018thd}, where $M_1$ and $M_2$ are either pseudoscalar~(P) or vector~(V) mesons. While PP $P$-wave production is compatible with experiment, the PV and VV modes are more likely to be $S$-wave production. More experimental data, especially for VV final states are needed. BESIII can measure a variety of these modes using the $\psi(2S)$ data sample, such as the PV modes $\tau \to \pi \rho \nu_{\tau}$, $\tau \to K \rho \nu_{\tau}$, $\tau \to \eta \rho \nu_{\tau}$, $\tau \to \phi K \nu_{\tau}$, and the VV modes $\tau \to \rho \rho \nu_{\tau}$, $\tau \to \rho \omega \nu_{\tau}$, $\tau \to K^{*} \rho \nu_{\tau}$.

\section{Exotic Decays and New Physics}
With the world's largest samples of $J/\psi$, $\psi$(2S), $D$, $D_s$, and $\Lambda_c$ decays at rest, plus large samples of $\eta$, $\omega$, $\eta’$, $\phi$, $\Lambda$, and $\chi_{cJ}$ decays, among others, BESIII has unique sensitivity to New Physics~(NP).  The scientific goals can be grouped into three categories. 
The first involves searching for unexpected deviations from the Standard Model~(SM) in precision measurements. 
This includes tests of lepton flavor universality, searches for weak decays of charmonia states, rare radiative decays of $D$/$D_s$ mesons, flavor changing neutral current processes, and measurements of $CP$ symmetry in decays of polarized quantum entangled hyperon pairs or in the triple-product of certain decays. 
Second, BESIII can search for SM-forbidden decays, such as baryon number violation in second generation quark decays, and baryon oscillations. In addition, lepton number violation can be probed in the decays of charmonia, light hadrons, and in charm meson decays with the same-charged leptons in the final states. 
Charged lepton flavor violation processes can also be investigated at BESIII in the decays of charm mesons, charmonia states, and the tau lepton, although the potential of the latter is less competitive compared to Belle II or future super tau-charm facilities. Third, BESIII data can be used to search for exotic and dark sector particles. Examples of the former include milli-charged particles, X17, and axion-like particles. Examples of the latter include low mass dark matter candidates in the invisible decays of charmonia and the charm mesons, dark photons, or the dark/light Higgs.

\subsection{Search for unexpected deviations from the SM }

As is well known, the decays of $\psi(nS)$ (n=1, 2) below the open-charm threshold are dominated by the strong or electromagnetic interactions where the intermediate gluons or virtual photons are produced by $c\bar{c}$ annihilation. However, flavor-changing weak decays of these states via virtual W bosons are also possible in the SM framework. For instance, the branching fractions of $J/\psi$ inclusive weak decays are estimated to be of the order of $10^{-10}$~\cite{Sanchis-Lozano:1993vyw}. 
As mentioned in Refs.~\cite{Datta:1998yq, Li:2012vk, Hill:1994hp}, the branching fractions of $J/\psi\to D(\bar{D})X$ (with $X$ denoting any hadron) can be enhanced by new interactions. NP models allow $\psi(nS)$ flavor-changing processes to occur with branching fractions around $10^{-6}$~\cite{Datta:1998yq}.
With the $10^{10}$ $J/\psi$ events, the current constraints of $J/\psi$ weak decays, including the hadronic and semi-leptonic weak decays, such as $J/\psi\to D^-_S\rho^+$~\cite{BESIII:2014xbo}, $J/\psi\to \bar{D}^0\bar{K}^{*0}$~\cite{BESIII:2014xbo}, $J/\psi\to D^{(*)-}_{S}e^+\nu_e+c.c.$~\cite{BESIII:2014pps} and $J/\psi\to D^0e^+e^-+c.c.$~\cite{BESIII:2017pez}, are expected to be improved by one or several orders of magnitude. These measurements will provide a more stringent experimental test of SM than previous searches, and hence further constrain the parameter space of NP models. These weak decays can also be searched for using the newly collected $3\times10^9$ $\psi(2S)$ events.

Rare radiative decays (such as $D\to\gamma\rho$) are most likely dominated by the long-distance (LD) SM contributions, which are quite difficult to compute. While there are opportunities to study NP effects in rare radiative transitions. 
The rare radiative decays of D mesons that are mediated by quark level FCNC transition $c\to \gamma u$ only proceed at the one loop level in SM. The absence of a super-heavy down-type quark in SM implies that GIM cancellation mechanism is very effective, making the charm sector of special interest for probing NP. The predicted short-distance (SD) contributions in SM of FCNC in the charm sector are well beyond the sensitivity of current experiments. Yet, some theoretical estimates suggest that the rates of FCNC processes could be enhanced by LD effects by several orders of magnitude.
Even though the charm production rate in $e^+e^-$ collisions near the charm threshold is lower than at hadron colliders and B-factories, BESIII has the benefit of lower multiplicity and the ability to impose powerful kinematic constraints, which can deliver good sensitivity. 

BESIII has the capability of testing $CP$ symmetry in hyperon decays, produced via $J/\psi \to B\bar B$, with $B\bar B$ denoting polarized, quantum-entangled hyperon pairs, which adds an exciting new dimension to the study of $CP$ violations.
We can exploit the triple-product asymmetry as a $CP$ violating observable. Studies of $CP$ violation in the hyperon decays using this approach were proposed in Ref. ~\cite{Bigi:2017eni}. The general direct $CP$ asymmetry only survives if there is a non-zero strong phase, whereas the triple-product asymmetry with the strong phase vanishes. Therefore the two methods are complementary, particularly if the strong phase is unknown. 

In recent years, hints of LFU violation have emerged in some semi-leptonic B-meson decays~\cite{BaBar:2012obs, BaBar:2013mob, Belle:2007qnm, Belle:2010tvu, LHCb:2015gmp, LHCb:2016aa, LHCb:2015aa, LHCb:2014vgu, Nomura:2014asa}. Theoretically, such violation could come from new lepton-flavor non-universal interactions~\cite{Altmannshofer:2017yso}, which might also be detectable in D-decays~\cite{Dorsner2017}. In Ref.~\cite{Fajfer:2015ixa}, it is argued that LFU violation may occur in $c \to s$ transitions due to an amplitude that includes a charged Higgs boson,  that arises in a two-Higgs-doublet model, interfering with the SM amplitude involving a $W^{\pm}$ boson. Therefore, it is important to test LFU with various decays. 
At BESIII, there have been many tests of LFU based on (semi-)leptonic decays of charm mesons, which agree with the SM predictions. However, these results are statistically limited. With expected 20 fb$^{-1}$ of data at 3.773 GeV, the precision of $R_{D^+}$ will be statistically limited to about 8\%. With 6 fb$^{-1}$ of data at 4.178 GeV, the precision of $R_{D^+_s}$ will be systematically limited to about 3\%.

\subsection{Search for SM-forbidden decays}
Some processes, while allowed by space-time symmetries, are forbidden in SM. Their observation, would constitute a high-impact discovery, as it would unambiguously point towards physics beyond SM. Examples include searches for the baryon-number violating transitions, the lepton-number violating decays, and the charged lepton-flavor violating (CLFV) decays.

The stability of ordinary matter implies baryon number ($B$) conservation. However, the observed fact of baryon asymmetry in the Universe shows that baryon number should be broken. There are many theoretical models in which
$B$ is not exactly conserved, with $B$ and lepton number ($L$) violated simultaneously while conserving $B-L$. There have been several searches for the $B$- or(and) $L$-violating processes at BESIII, such as $J/\psi\to\Lambda_c^+e^-+c.c.$\cite{BESIII:2018cls}, $D\to K \pi e^+ e^+$\cite{BESIII:2019oef}, $D^+\to\bar\Lambda(\bar\Sigma^0)e^+$ and $D^+\to\Lambda(\Sigma^0)e^+$\cite{BESIII:2019udi}, $D^0 \rightarrow \overline{p} e^+$ and $D^0 \rightarrow pe^-$\cite{BESIII:2021krj}.  All such processes will be probed further and more processes will be searched for.  Hyperon samples also could be used for such searches. The lepton-number-violating $\Sigma^-$ decays was already searched for with $1.3 \times 10^{9}$ $J/\psi$ events~\cite{BESIII:2020iwk}. Given that such searches are almost background-free, one would expect the bounds from the $10^{10}$ $J/\psi$ sample by scaling  with the statistics, as detailed in Ref.~\cite{Li:2016tlt}. Full invisible decays of hyperons are also within the reach with BESIII. For example, $\Lambda$ invisible decay is searched for the first time, and its branching ratio is found to be less than $7.4 \times 10^{-5}$ at $90\%$ CL with the $J/\psi$ sample~\cite{BESIII:2021slv}.

The $\Delta B = 2$  baryon-antibaryon oscillation is another crucial test of baryon number violation, as pointed out long ago~\cite{Mohapatra:1980qe}.  There was a proposal to search for
$\Lambda$-$\bar\Lambda$ oscillations in the decay of $J/\psi\to\Lambda\bar\Lambda$~\cite{Kang:2009xt}. At BESIII, the decays of $J/\psi\to pK^-\bar\Lambda$ and $J/\psi\to \Lambda \bar\Lambda$ have  very simple
final states and are almost background free,  thus well suited for searching for $\Lambda-\bar\Lambda$ oscillations.  The probability of generating a
final $\Lambda$ from an initial $\bar\Lambda$ can be determined by the ratio of wrong sign events over right sign events. With the data sample of $10^{10}$ $J/\psi$
events, the upper limit of the $\Lambda$-$\bar\Lambda$ oscillation rate will be at $10^{-6}$ level and the constraint on
$\delta m_{\Lambda\bar\Lambda}$ will be reduced to $10^{-16}$ MeV. A time-dependent analysis of the produced $\Lambda$-$\bar\Lambda$ pairs from the $J/\psi$ decays could also be investigated, taking advantage of their long mean flight distance  in the detector.  

CLFV processes are highly suppressed in the SM by finite but tiny neutrino masses, so their branching
fractions are predicted to be negligibly small. Yet, various theoretical models predict the rates for CLFV transitions to be large enough
to be experimentally observable and would provide direct evidence for NP. In the charmonium system,  by
analyzing a data sample of 58 million $J/\psi$ events, BESII obtained a limit ${\cal B}(J/\psi \to e \mu ) < 1.1 \times 10^{-6}$~\cite{BES:2003zru}, ${\cal B}(J/\psi \to \mu \tau) < 2.0 \times 10^{-6}$ and ${\cal B}(J/\psi \to e \tau) < 8.3 \times 10^{-6}$~\cite{BES:2004jiw}.  BESIII set the upper limit of ${\cal B}(J/\psi \to e \mu ) < 1.6 \times 10^{-7}$ at the 90\%
C.L.~\cite{BESIII:2013jau} with 255 million $J/\psi$ decays.  Recently, with the $10^{10}$ $J/\psi$ datasets, BESIII performed the search 
  $J/\psi\to e^{\pm}\tau^{\mp}$ with $\tau^{\mp} \to \pi^{\mp}\pi^0\nu_{\tau}$ and set the upper limit   $\mathcal{B}(J/\psi\to e^{\pm}\tau^{\mp})<7.5\times10^{-8}$ at the 90$\%$ C.L., which improves the previous limit by two orders of magnitude.  Other channels are expected to be updated with similar magnitude improvement and new channels will be probed.

CLFV and lepton-number-violating (LNV) processes can also be probed in $D$-meson decays at BESIII. No evidence has been found for the
$D$-meson decays with either CLFV or LNV. The present experimental bounds on the branching fractions are generally set at the level of
$10^{-6}$ to $10^{-5}$ (with a notable exception of $D^0\to \mu e$, where ${\cal B} (D^0\to \mu^\pm e^\mp) < 1.3 \times 10^{-8}$)~\cite{LHCb:2015pce}. $D$ decays with LNV, such as $D^+ \to l^+ l^+ X^-$ and $D_s^+ \to l^+ l^+ X^-$, are also forbidden in the minimal SM, but are possible if massive neutrinos are Majorana particles.  These could be probed with the incoming 20 $fb^{-1}$~$\psi(3773)$ dataset, due to the clean environment and low charge confusion rates.

% RM EDITED: 4/18
\subsection{Search for exotic and dark sectors}
%milli-charged
Electric charge quantization is related to the existence of the magnetic monopole, which, despite many searches, has not been established by experiment. 
As an alternative,
theoretical models have been developed in recent years that predict new particles, not in the Standard Model,
that carry very small electric charges, referred to as millicharged particles. These are not easily detectable by experiment. 
A sensitive search for millicharged particles at an electron-positron experiment such as BESIII is possible by looking for missing energy in $e^+e^-$ collisions. By reconstructing a single photon and searching for missing millicharged particles using energy-momentum conservation, BESIII can probe a $\cal{O}$(GeV) millicharged particle with charge $\epsilon\sim10^{-3}$ based on the current data set~\cite{Liu:2018jdi}. A more detailed analysis in Ref.~\cite{Liang:2019zkb} shows with 17~ fb$^{-1}$ data at BESIII, the sensitivity for a millicharged particle with mass 0.1 – 1.5 GeV can even achieve $\epsilon<10^{-3}$, which is competitive to Belle II in the low mass region.

The X17 particle, recently reported in a nuclear physics experiment, was once considered the carrier of a fifth force, beyond the four elementary forces in the Standard Model. The X17 was first reported by analyzing angular correlations of $e^+e^-$ pairs originating from $^8$Be M1 transitions~\cite{Krasznahorkay:2015iga}, and was further confirmed in a $^4$He M0 transition~\cite{Krasznahorkay:2019lyl}  by the Atomki group. Various attempts have been made to identify the X17 signal, for example as a QCD axion, a light $Z_0$, or a light pseudoscalar particle. 
As long as the X17 candidate exists, it is important to extend the searches to $e^+e^-$ experiments, such as BESIII. Since the X17 candidate decays to $e^+e^-$, it can be searched for at BESIII in the process $e^+e^-\to\gamma {\rm X17}\to\gamma e^+e^-$~\cite{Jiang:2018jqp}. Other decays, such as $J/\psi\to\gamma {\rm X17}\to \gamma \mu^+\mu^-$, $\psi(2S)\to {\rm X17}\chi_{cJ}$~($J=0,1,2$), and $\pi^0\to\gamma {\rm X17}\to\gamma e^+e^-$ are also ideal places to hunt for the X17 at BESIII. Depending on the source of X17 production, the sensitivity to the decay branching fraction can reach levels of $10^{-5}$ - $10^{-9}$.

Axions and axion-like particles ($a$) are pseudoscalar particles that were first introduced to solve the strong CP problem. 
Intensive searches for axions in various energy regions are currently ongoing.
At BESIII, it is most promising to search for axion candidates with mass $\cal{O}$(GeV), and if the results are negative, stringent constraints on the axion-photon coupling can be set. Compared to the Belle II experiment, which searches for axions in the process $e^+e^-\to\gamma a\to\gamma\gamma\gamma$~\cite{Belle-II:2020jti}, BESIII can search for axions using the world's largest samples of $J/\psi$ and $\psi(2S)$ decays. The search is already ongoing, and we expect to achieve a better sensitivity in the 1-3 GeV mass region than the current Belle II limit.

A dark photon is a hypothetical boson which is often taken to be a mediator associated with an extra U(1) symmetry. It serves as a portal between dark matter and Standard Model particles. Searches for dark photon candidates have been performed in a wide mass region. At BESIII, the interesting mass region for a possible dark photon candidate is $\cal{O}$(GeV). Using 2.93~fb$^{-1}$ of data taken at $\sqrt{s}=3.773$~GeV, BESIII searched for $\gamma^\prime\to e^+e^-/\mu^+\mu^-$ and reported negative results~\cite{BESIII:2017fwv}. The obtained exclusion limit is a bit looser than BABAR~\cite{BaBar:2014zli}. With an additional 17~fb$^{-1}$ of data, we will achieve better sensitivity in the mass region between 1.5 and 3.4 GeV, and we expect to improve upon the BABAR limit. Dark photons with invisible decays are also studied at BESIII using single photon signals. With the single photon trigger active, we also expect to achieve a better sensitivity than BABAR in this mass region.

A dark baryon candidate was recently proposed in hyperon decays~\cite{Alonso-Alvarez:2021oaj}. 
With 10 billion $J/\psi$ decays, and considering the 
decay branch fractions of $J/\psi$ to hyperon pairs, such as $\Lambda$, $\Xi$, etc., are on the order of $10^{-3}$,
BESIII has access to millions of hyperon pairs.
With such large hyperon samples, the dark baryon candidate can be searched for in $\Lambda$, $\Xi$, and $\Sigma$ decays, accompanied by a $\phi$, $\pi^0$, or $\gamma$. A recent search for $\Lambda\to\gamma$+invisible has been performed by BESIII~\cite{BESIII:2021slv}, with negative results reported. The upper limit for the branching fraction is $7.4\times 10^{-5}$. We expect similar sensitivity can be achieved in $\Sigma$ and $\Xi$ decays.

Dark sector and light Higgs candidates with invisible decays can also be searched for in charmonium decays, with a typical branching fraction sensitivity from $10^{-6}$ to $10^{-9}$. In addition, BESIII has a large $\eta^\prime$ data sample produced from $J/\psi$ decays (even larger than KLOE). Thus, dark sectors with invisible or visible decays can also been searched for using $\eta^\prime$ decays, and the expected sensitivity is on the $\cal{O}$($10^{-5}$) level.

\bibliography{BESIIISnowMassWP}{}

\providecommand{\href}[2]{#2}\begingroup\raggedright\begin{thebibliography}{100}

\bibitem{whitepaper}
{\scshape BESIII} collaboration, \emph{{Future Physics Programme of BESIII}},
  \href{https://doi.org/10.1088/1674-1137/44/4/040001}{\emph{Chin. Phys. C}
  {\bfseries 44} (2020) 040001}
  [\href{https://arxiv.org/abs/1912.05983}{{\ttfamily 1912.05983}}].

\bibitem{gampipi}
{\scshape BESIII} collaboration, \emph{{Amplitude analysis of the
  \ensuremath{\pi}$^0$\ensuremath{\pi}$^0$ system produced in radiative
  J/\ensuremath{\psi} decays}},
  \href{https://doi.org/10.1103/PhysRevD.92.052003}{\emph{Phys. Rev. D}
  {\bfseries 92} (2015) 052003}
  [\href{https://arxiv.org/abs/1506.00546}{{\ttfamily 1506.00546}}].

\bibitem{gamKsKs}
{\scshape BESIII} collaboration, \emph{{Amplitude analysis of the $K_{S}K_{S}$
  system produced in radiative $J/\psi$ decays}},
  \href{https://doi.org/10.1103/PhysRevD.98.072003}{\emph{Phys. Rev. D}
  {\bfseries 98} (2018) 072003}
  [\href{https://arxiv.org/abs/1808.06946}{{\ttfamily 1808.06946}}].

\bibitem{gametaeta}
{\scshape BESIII} collaboration, \emph{{Partial wave analysis of $J/\psi \to
  \gamma \eta \eta$}},
  \href{https://doi.org/10.1103/PhysRevD.87.092009}{\emph{Phys. Rev. D}
  {\bfseries 87} (2013) 092009}
  [\href{https://arxiv.org/abs/1301.0053}{{\ttfamily 1301.0053}}].

\bibitem{gamphiphi}
{\scshape BESIII} collaboration, \emph{{Observation of pseudoscalar and tensor
  resonances in $J/\psi\to \gamma \phi \phi$}},
  \href{https://doi.org/10.1103/PhysRevD.93.112011}{\emph{Phys. Rev. D}
  {\bfseries 93} (2016) 112011}
  [\href{https://arxiv.org/abs/1602.01523}{{\ttfamily 1602.01523}}].

\bibitem{bonn}
A.V.~Sarantsev, I.~Denisenko, U.~Thoma and E.~Klempt, \emph{{Scalar isoscalar
  mesons and the scalar glueball from radiative $J/\psi$ decays}},
  \href{https://doi.org/10.1016/j.physletb.2021.136227}{\emph{Phys. Lett. B}
  {\bfseries 816} (2021) 136227}
  [\href{https://arxiv.org/abs/2103.09680}{{\ttfamily 2103.09680}}].

\bibitem{jpac}
{\scshape Joint Physics Analysis Center} collaboration, \emph{{Scalar and
  tensor resonances in $J/\psi $ radiative decays}},
  \href{https://doi.org/10.1140/epjc/s10052-022-10014-8}{\emph{Eur. Phys. J. C}
  {\bfseries 82} (2022) 80} [\href{https://arxiv.org/abs/2110.00027}{{\ttfamily
  2110.00027}}].

\bibitem{X1835}
{\scshape BESIII} collaboration, \emph{{Observation of an anomalous line shape
  of the $\eta^{\prime}\pi^{+}\pi^{-}$ mass spectrum near the $p\bar{p}$ mass
  threshold in $J/\psi\rightarrow\gamma\eta^{\prime}\pi^{+}\pi^{-}$}},
  \href{https://doi.org/10.1103/PhysRevLett.117.042002}{\emph{Phys. Rev. Lett.}
  {\bfseries 117} (2016) 042002}
  [\href{https://arxiv.org/abs/1603.09653}{{\ttfamily 1603.09653}}].

\bibitem{pi1chic}
{\scshape BESIII} collaboration, \emph{{Amplitude analysis of the $\chi_{c1}
  \to \eta\pi^+\pi^-$ decays}},
  \href{https://doi.org/10.1103/PhysRevD.95.032002}{\emph{Phys. Rev. D}
  {\bfseries 95} (2017) 032002}
  [\href{https://arxiv.org/abs/1610.02479}{{\ttfamily 1610.02479}}].

\bibitem{eta11}
{\scshape BESIII} collaboration, \emph{{Observation of an isoscalar resonance
  with exotic $J^{PC}=1^{-+}$ quantum numbers in
  $J/\psi\rightarrow\gamma\eta\eta'$}},
  \href{https://arxiv.org/abs/2202.00621}{{\ttfamily 2202.00621}}.

\bibitem{eta12}
{\scshape BESIII} collaboration, \emph{{Partial wave analysis of
  $J/\psi\rightarrow\gamma\eta\eta'$}},
  \href{https://arxiv.org/abs/2202.00623}{{\ttfamily 2202.00623}}.

\bibitem{mixing1}
{\scshape BESIII} collaboration, \emph{{Study of $a_0^0(980) - f_0(980)$
  mixing}}, \href{https://doi.org/10.1103/PhysRevD.83.032003}{\emph{Phys. Rev.
  D} {\bfseries 83} (2011) 032003}
  [\href{https://arxiv.org/abs/1012.5131}{{\ttfamily 1012.5131}}].

\bibitem{mixing2}
{\scshape BESIII} collaboration, \emph{{Observation of
  $a^{0}_{0}(980)$-$f_{0}(980)$ mixing}},
  \href{https://doi.org/10.1103/PhysRevLett.121.022001}{\emph{Phys. Rev. Lett.}
  {\bfseries 121} (2018) 022001}
  [\href{https://arxiv.org/abs/1802.00583}{{\ttfamily 1802.00583}}].

\bibitem{RevModPhys.91.015003}
S.D.~Bass and P.~Moskal, \emph{${\ensuremath{\eta}}^{\ensuremath{'}}$ and
  $\ensuremath{\eta}$ mesons with connection to anomalous glue},
  \href{https://doi.org/10.1103/RevModPhys.91.015003}{\emph{Rev. Mod. Phys.}
  {\bfseries 91} (2019) 015003}.

\bibitem{PhysRevLett.118.012001}
{\scshape BESIII} collaboration, \emph{Amplitude analysis of the decays
  ${\ensuremath{\eta}}^{\ensuremath{'}}\ensuremath{\rightarrow}{\ensuremath{\pi}}^{+}{\ensuremath{\pi}}^{\ensuremath{-}}{\ensuremath{\pi}}^{0}$
  and
  ${\ensuremath{\eta}}^{\ensuremath{'}}\ensuremath{\rightarrow}{\ensuremath{\pi}}^{0}{\ensuremath{\pi}}^{0}{\ensuremath{\pi}}^{0}$},
  \href{https://doi.org/10.1103/PhysRevLett.118.012001}{\emph{Phys. Rev. Lett.}
  {\bfseries 118} (2017) 012001}.

\bibitem{PhysRevD.92.012001}
{\scshape BESIII} collaboration, \emph{Observation of the dalitz decay
  $\ensuremath{\eta}\ensuremath{'}\ensuremath{\rightarrow}\ensuremath{\gamma}{e}^{+}{e}^{\ensuremath{-}}$},
  \href{https://doi.org/10.1103/PhysRevD.92.012001}{\emph{Phys. Rev. D}
  {\bfseries 92} (2015) 012001}.

\bibitem{PhysRevLett.112.251801}
{\scshape BESIII} collaboration, \emph{Observation of
  $\ensuremath{\eta}\ensuremath{'}\ensuremath{\rightarrow}{\ensuremath{\pi}}^{+}{\ensuremath{\pi}}^{\ensuremath{-}}{\ensuremath{\pi}}^{+}{\ensuremath{\pi}}^{\ensuremath{-}}$
  and
  $\ensuremath{\eta}\ensuremath{'}\ensuremath{\rightarrow}{\ensuremath{\pi}}^{+}{\ensuremath{\pi}}^{\ensuremath{-}}{\ensuremath{\pi}}^{0}{\ensuremath{\pi}}^{0}$},
  \href{https://doi.org/10.1103/PhysRevLett.112.251801}{\emph{Phys. Rev. Lett.}
  {\bfseries 112} (2014) 251801}.

\bibitem{PhysRevLett.113.039903}
{\scshape BESIII} collaboration, \emph{Publisher's note: Observation of
  $\ensuremath{\eta}\ensuremath{'}\ensuremath{\rightarrow}{\ensuremath{\pi}}^{+}{\ensuremath{\pi}}^{\ensuremath{-}}{\ensuremath{\pi}}^{+}{\ensuremath{\pi}}^{\ensuremath{-}}$and
  $\ensuremath{\eta}\ensuremath{'}\ensuremath{\rightarrow}{\ensuremath{\pi}}^{+}{\ensuremath{\pi}}^{\ensuremath{-}}{\ensuremath{\pi}}^{0}{\ensuremath{\pi}}^{0}$
  [phys. rev. lett. 112, 251801 (2014)]},
  \href{https://doi.org/10.1103/PhysRevLett.113.039903}{\emph{Phys. Rev. Lett.}
  {\bfseries 113} (2014) 039903}.

\bibitem{PhysRevLett.122.142002}
{\scshape BESIII} collaboration, \emph{Precision measurement of the branching
  fractions of ${\ensuremath{\eta}}^{\ensuremath{'}}$ decays},
  \href{https://doi.org/10.1103/PhysRevLett.122.142002}{\emph{Phys. Rev. Lett.}
  {\bfseries 122} (2019) 142002}.

\bibitem{Escribano_2018}
R.~Escribano and S.~Gonz{\`{a}}lez-Sol{\'{\i}}s, \emph{A data-driven approach
  to $\pi^0\, \eta$ and $\eta'$ single and double dalitz decays},
  \href{https://doi.org/10.1088/1674-1137/42/2/023109}{\emph{Chinese Physics C}
  {\bfseries 42} (2018) 023109}.

\bibitem{petri2010anomalous}
T.~Petri, \emph{Anomalous decays of pseudoscalar mesons},
  \href{https://arxiv.org/abs/1010.2378}{{\ttfamily 1010.2378}}.

\bibitem{refId0}
{Kubis, Bastian} and {Schneider, Sebastian P.}, \emph{The cusp effect in
  $\eta^\prime \to \eta \pi \pi$ decays},
  \href{https://doi.org/10.1140/epjc/s10052-009-1054-7}{\emph{Eur. Phys. J. C}
  {\bfseries 62} (2009) 511}.

\bibitem{PhysRevD.104.092004}
{\scshape BESIII} collaboration, \emph{Measurement of the absolute branching
  fractions of
  $\ensuremath{J}/\ensuremath{\psi}\ensuremath{\rightarrow}\ensuremath{\gamma}\ensuremath{\eta}$
  and $\ensuremath{\eta}$ decay modes},
  \href{https://doi.org/10.1103/PhysRevD.104.092004}{\emph{Phys. Rev. D}
  {\bfseries 104} (2021) 092004}.

\bibitem{PhysRevLett.108.182001}
{\scshape BESIII} collaboration, \emph{First observation of
  $\ensuremath{\eta}(1405)$ decays into ${f}_{0}(980){\ensuremath{\pi}}^{0}$},
  \href{https://doi.org/10.1103/PhysRevLett.108.182001}{\emph{Phys. Rev. Lett.}
  {\bfseries 108} (2012) 182001}.

\bibitem{PhysRevD.91.052017}
{\scshape BESIII} collaboration, \emph{{Study of
  $J/\ensuremath{\psi}\ensuremath{\rightarrow}\ensuremath{\eta}\ensuremath{\phi}{\ensuremath{\pi}}^{+}{\ensuremath{\pi}}^{\ensuremath{-}}$
  at BESIII}}, \href{https://doi.org/10.1103/PhysRevD.91.052017}{\emph{Phys.
  Rev. D} {\bfseries 91} (2015) 052017}.

\bibitem{PhysRevD.98.112007}
{\scshape BESIII} collaboration, \emph{Dalitz plot analysis of the decay
  $\ensuremath{\omega}\ensuremath{\rightarrow}{\ensuremath{\pi}}^{+}{\ensuremath{\pi}}^{\ensuremath{-}}{\ensuremath{\pi}}^{0}$},
  \href{https://doi.org/10.1103/PhysRevD.98.112007}{\emph{Phys. Rev. D}
  {\bfseries 98} (2018) 112007}.

\bibitem{Capstick:2000qj}
S.~Capstick and W.~Roberts, \emph{{Quark models of baryon masses and decays}},
  \href{https://doi.org/10.1016/S0146-6410(00)00109-5}{\emph{Prog. Part. Nucl.
  Phys.} {\bfseries 45} (2000) S241}
  [\href{https://arxiv.org/abs/nucl-th/0008028}{{\ttfamily nucl-th/0008028}}].

\bibitem{BES:2001gvq}
{\scshape BES} collaboration, \emph{{Study of N* production from $J / \psi \to
  p \bar{p} \eta$}},
  \href{https://doi.org/10.1016/S0370-2693(01)00605-0}{\emph{Phys. Lett. B}
  {\bfseries 510} (2001) 75}
  [\href{https://arxiv.org/abs/hep-ex/0105011}{{\ttfamily hep-ex/0105011}}].

\bibitem{BES:2004gwe}
{\scshape BES} collaboration, \emph{{Observation of two new N* peaks in $J/\psi
  \to p \pi^- \bar{n}$ and $\bar{p} \pi^+ n$ decays}},
  \href{https://doi.org/10.1103/PhysRevLett.97.062001}{\emph{Phys. Rev. Lett.}
  {\bfseries 97} (2006) 062001}
  [\href{https://arxiv.org/abs/hep-ex/0405030}{{\ttfamily hep-ex/0405030}}].

\bibitem{BES:2009ufh}
{\scshape BES} collaboration, \emph{{Partial wave analysis of $J/\psi \to p
  \bar{p} \pi^0$}},
  \href{https://doi.org/10.1103/PhysRevD.80.052004}{\emph{Phys. Rev. D}
  {\bfseries 80} (2009) 052004}
  [\href{https://arxiv.org/abs/0905.1562}{{\ttfamily 0905.1562}}].

\bibitem{BESIII:2013xkm}
{\scshape BESIII} collaboration, \emph{{Partial wave analysis of
  \ensuremath{\psi}(2S) \textrightarrow{} $p\overline{p}\eta$}},
  \href{https://doi.org/10.1103/PhysRevD.88.032010}{\emph{Phys. Rev. D}
  {\bfseries 88} (2013) 032010}
  [\href{https://arxiv.org/abs/1304.1973}{{\ttfamily 1304.1973}}].

\bibitem{BESIII:2012ssm}
{\scshape BESIII} collaboration, \emph{{Observation of two new $N^*$ resonances
  in the decay $\psi(3686) \rightarrow p\bar{p}\pi^0$}},
  \href{https://doi.org/10.1103/PhysRevLett.110.022001}{\emph{Phys. Rev. Lett.}
  {\bfseries 110} (2013) 022001}
  [\href{https://arxiv.org/abs/1207.0223}{{\ttfamily 1207.0223}}].

\bibitem{BESIII:2013bgg}
{\scshape BESIII} collaboration, \emph{{Observation of the decay $\psi(3686)$
  $\rightarrow$ $\Lambda\bar\Sigma^{\pm}\pi^{\mp}+c.c.$}},
  \href{https://doi.org/10.1103/PhysRevD.88.112007}{\emph{Phys. Rev. D}
  {\bfseries 88} (2013) 112007}
  [\href{https://arxiv.org/abs/1310.5826}{{\ttfamily 1310.5826}}].

\bibitem{Benmerrouche:1996ij}
M.~Benmerrouche, N.C.~Mukhopadhyay and J.F.~Zhang, \emph{{Model independent
  extraction of the N* (1535) electrostrong form-factor from eta
  electroproduction}},
  \href{https://doi.org/10.1103/PhysRevLett.77.4716}{\emph{Phys. Rev. Lett.}
  {\bfseries 77} (1996) 4716}
  [\href{https://arxiv.org/abs/nucl-th/9611032}{{\ttfamily nucl-th/9611032}}].

\bibitem{Olsson:1976st}
M.G.~Olsson and E.T.~Osypowski, \emph{{Vector Mesonic and Unitarity Effects in
  Low-Energy Photoproduction}},
  \href{https://doi.org/10.1103/PhysRevD.17.174}{\emph{Phys. Rev. D} {\bfseries
  17} (1978) 174}.

\bibitem{Wang:1993dwa}
J.-X.~Wang, \emph{{Automatic calculation of Feynman loop diagram. 1. Generation
  of simplified form of amplitude}},
  \href{https://doi.org/10.1016/0010-4655(93)90010-A}{\emph{Comput. Phys.
  Commun.} {\bfseries 77} (1993) 263}.

\bibitem{BESIII:2018cnd}
{\scshape BESIII} collaboration, \emph{{Polarization and Entanglement in
  Baryon-Antibaryon Pair Production in Electron-Positron Annihilation}},
  \href{https://doi.org/10.1038/s41567-019-0494-8}{\emph{Nature Phys.}
  {\bfseries 15} (2019) 631}
  [\href{https://arxiv.org/abs/1808.08917}{{\ttfamily 1808.08917}}].

\bibitem{Chao:1996sf}
K.-T.~Chao, Y.-F.~Gu and S.F.~Tuan, \emph{{Gluonia and charmonium decays}},
  {\emph{Commun. Theor. Phys.} {\bfseries 25} (1996) 471}.

\bibitem{Mo:2006cy}
X.-H.~Mo, C.-Z.~Yuan and P.~Wang, \emph{{Study of the Rho-pi Puzzle in
  Charmonium Decays}}, {\emph{Chin. Phys. C} {\bfseries 31} (2007) 686}
  [\href{https://arxiv.org/abs/hep-ph/0611214}{{\ttfamily hep-ph/0611214}}].

\bibitem{Wang:2021dxw}
H.~Wang and C.-Z.~Yuan, \emph{{New puzzle in charmonium decays}},
  \href{https://arxiv.org/abs/2112.08584}{{\ttfamily 2112.08584}}.

\bibitem{BESIII:2015iqd}
{\scshape BESIII} collaboration, \emph{{Observation of the $\psi(1^3D_2)$ state
  in $e^+e^-\to\pi^+\pi^-\gamma\chi_{c1}$ at BESIII}},
  \href{https://doi.org/10.1103/PhysRevLett.115.011803}{\emph{Phys. Rev. Lett.}
  {\bfseries 115} (2015) 011803}
  [\href{https://arxiv.org/abs/1503.08203}{{\ttfamily 1503.08203}}].

\bibitem{BESIII:2021yvc}
{\scshape BESIII} collaboration, \emph{{Cross section measurements of the
  $e^+e^-\to D^{*+}D^{*-}$ and $e^+e^-\to D^{*+}D^{-}$ processes at
  center-of-mass energies from 4.085 to 4.600 GeV}},
  \href{https://arxiv.org/abs/2112.06477}{{\ttfamily 2112.06477}}.

\bibitem{BESIII:2020bgb}
{\scshape BESIII} collaboration, \emph{{Observation of the $Y(4220)$ and
  $Y(4360)$ in the process $e^{+}e^{-} \to \eta J/\psi$}},
  \href{https://doi.org/10.1103/PhysRevD.102.031101}{\emph{Phys. Rev. D}
  {\bfseries 102} (2020) 031101}
  [\href{https://arxiv.org/abs/2003.03705}{{\ttfamily 2003.03705}}].

\bibitem{BESIII:2019tdo}
{\scshape BESIII} collaboration, \emph{{Observation of $e^{+}e^{-}\rightarrow
  \pi^{+}\pi^{-}\psi(3770)$ and $D_{1}(2420)^{0}\bar{D}^{0} + c.c.$}},
  \href{https://doi.org/10.1103/PhysRevD.100.032005}{\emph{Phys. Rev. D}
  {\bfseries 100} (2019) 032005}
  [\href{https://arxiv.org/abs/1903.08126}{{\ttfamily 1903.08126}}].

\bibitem{BESIII:2022yga}
{\scshape BESIII} collaboration, \emph{{Observation of resonance structures in
  $e^+e^-\to \pi^+\pi^-\psi_2(3823)$ and mass measurement of $\psi_2(3823)$}},
  \href{https://arxiv.org/abs/2203.05815}{{\ttfamily 2203.05815}}.

\bibitem{BESIII:2021yal}
{\scshape BESIII} collaboration, \emph{{Measurement of $e^+e^-\to\gamma
  \chi_{c0,c1,c2}$ cross sections at center-of-mass energies between 3.77 and
  4.60~GeV}}, \href{https://doi.org/10.1103/PhysRevD.104.092001}{\emph{Phys.
  Rev. D} {\bfseries 104} (2021) 092001}
  [\href{https://arxiv.org/abs/2107.03604}{{\ttfamily 2107.03604}}].

\bibitem{BESIII:2021qmo}
{\scshape BESIII} collaboration, \emph{{Search for new decay modes of the
  $\psi_2(3823)$ and the process $e^+e^-\rightarrow\pi^0\pi^0\psi_2(3823)$}},
  \href{https://doi.org/10.1103/PhysRevD.103.L091102}{\emph{Phys. Rev. D}
  {\bfseries 103} (2021) L091102}
  [\href{https://arxiv.org/abs/2102.10845}{{\ttfamily 2102.10845}}].

\bibitem{Lebed:2016hpi}
R.F.~Lebed, R.E.~Mitchell and E.S.~Swanson, \emph{{Heavy-Quark QCD Exotica}},
  \href{https://doi.org/10.1016/j.ppnp.2016.11.003}{\emph{Prog. Part. Nucl.
  Phys.} {\bfseries 93} (2017) 143}
  [\href{https://arxiv.org/abs/1610.04528}{{\ttfamily 1610.04528}}].

\bibitem{Chen:2016qju}
H.-X.~Chen, W.~Chen, X.~Liu and S.-L.~Zhu, \emph{{The hidden-charm pentaquark
  and tetraquark states}},
  \href{https://doi.org/10.1016/j.physrep.2016.05.004}{\emph{Phys. Rept.}
  {\bfseries 639} (2016) 1} [\href{https://arxiv.org/abs/1601.02092}{{\ttfamily
  1601.02092}}].

\bibitem{Esposito:2016noz}
A.~Esposito, A.~Pilloni and A.D.~Polosa, \emph{{Multiquark Resonances}},
  \href{https://doi.org/10.1016/j.physrep.2016.11.002}{\emph{Phys. Rept.}
  {\bfseries 668} (2017) 1} [\href{https://arxiv.org/abs/1611.07920}{{\ttfamily
  1611.07920}}].

\bibitem{Guo:2017jvc}
F.-K.~Guo, C.~Hanhart, U.-G.~Mei\ss{}ner, Q.~Wang, Q.~Zhao and B.-S.~Zou,
  \emph{{Hadronic molecules}},
  \href{https://doi.org/10.1103/RevModPhys.90.015004}{\emph{Rev. Mod. Phys.}
  {\bfseries 90} (2018) 015004}
  [\href{https://arxiv.org/abs/1705.00141}{{\ttfamily 1705.00141}}].

\bibitem{Ali:2017jda}
A.~Ali, J.S.~Lange and S.~Stone, \emph{{Exotics: Heavy Pentaquarks and
  Tetraquarks}}, \href{https://doi.org/10.1016/j.ppnp.2017.08.003}{\emph{Prog.
  Part. Nucl. Phys.} {\bfseries 97} (2017) 123}
  [\href{https://arxiv.org/abs/1706.00610}{{\ttfamily 1706.00610}}].

\bibitem{Olsen:2017bmm}
S.L.~Olsen, T.~Skwarnicki and D.~Zieminska, \emph{{Nonstandard heavy mesons and
  baryons: Experimental evidence}},
  \href{https://doi.org/10.1103/RevModPhys.90.015003}{\emph{Rev. Mod. Phys.}
  {\bfseries 90} (2018) 015003}
  [\href{https://arxiv.org/abs/1708.04012}{{\ttfamily 1708.04012}}].

\bibitem{Brambilla:2019esw}
N.~Brambilla, S.~Eidelman, C.~Hanhart, A.~Nefediev, C.-P.~Shen, C.E.~Thomas
  et~al., \emph{{The $XYZ$ states: experimental and theoretical status and
  perspectives}},
  \href{https://doi.org/10.1016/j.physrep.2020.05.001}{\emph{Phys. Rept.}
  {\bfseries 873} (2020) 1} [\href{https://arxiv.org/abs/1907.07583}{{\ttfamily
  1907.07583}}].

\bibitem{Yuan:2021wpg}
C.-Z.~Yuan, \emph{{Charmonium and charmoniumlike states at the BESIII
  experiment}}, \href{https://doi.org/10.1093/nsr/nwab182}{\emph{Natl. Sci.
  Rev.} {\bfseries 8} (2021) nwab182}
  [\href{https://arxiv.org/abs/2102.12044}{{\ttfamily 2102.12044}}].

\bibitem{BESIII:2016bnd}
{\scshape BESIII} collaboration, \emph{{Precise measurement of the $e^+e^-\to
  \pi^+\pi^-J/\psi$ cross section at center-of-mass energies from 3.77 to 4.60
  GeV}}, \href{https://doi.org/10.1103/PhysRevLett.118.092001}{\emph{Phys. Rev.
  Lett.} {\bfseries 118} (2017) 092001}
  [\href{https://arxiv.org/abs/1611.01317}{{\ttfamily 1611.01317}}].

\bibitem{BESIII:2016adj}
{\scshape BESIII} collaboration, \emph{{Evidence of Two Resonant Structures in
  $e^+ e^- \to \pi^+ \pi^- h_c$}},
  \href{https://doi.org/10.1103/PhysRevLett.118.092002}{\emph{Phys. Rev. Lett.}
  {\bfseries 118} (2017) 092002}
  [\href{https://arxiv.org/abs/1610.07044}{{\ttfamily 1610.07044}}].

\bibitem{BESIII:2014rja}
{\scshape BESIII} collaboration, \emph{{Study of $e^+e^-\to\omega\chi_{cJ}$ at
  center-of-mass energies from 4.21 to 4.42 GeV}},
  \href{https://doi.org/10.1103/PhysRevLett.114.092003}{\emph{Phys. Rev. Lett.}
  {\bfseries 114} (2015) 092003}
  [\href{https://arxiv.org/abs/1410.6538}{{\ttfamily 1410.6538}}].

\bibitem{BESIII:2018iea}
{\scshape BESIII} collaboration, \emph{{Evidence of a resonant structure in the
  $e^+e^-\to \pi^+D^0D^{*-}$ cross section between 4.05 and 4.60 GeV}},
  \href{https://doi.org/10.1103/PhysRevLett.122.102002}{\emph{Phys. Rev. Lett.}
  {\bfseries 122} (2019) 102002}
  [\href{https://arxiv.org/abs/1808.02847}{{\ttfamily 1808.02847}}].

\bibitem{BESIII:2017tqk}
{\scshape BESIII} collaboration, \emph{{Measurement of $e^{+}e^{-}\rightarrow
  \pi^{+}\pi^{-}\psi(3686)$ from 4.008 to 4.600\textasciitilde{}GeV and
  observation of a charged structure in the $\pi^{\pm}\psi(3686)$ mass
  spectrum}}, \href{https://doi.org/10.1103/PhysRevD.96.032004}{\emph{Phys.
  Rev. D} {\bfseries 96} (2017) 032004}
  [\href{https://arxiv.org/abs/1703.08787}{{\ttfamily 1703.08787}}].

\bibitem{BESIII:2021njb}
{\scshape BESIII} collaboration, \emph{{Cross section measurement of
  $e^+e^-\rightarrow\pi^+\pi^-(3686)$ from $\sqrt{S}=4.0076$ to 4.6984~GeV}},
  \href{https://doi.org/10.1103/PhysRevD.104.052012}{\emph{Phys. Rev. D}
  {\bfseries 104} (2021) 052012}
  [\href{https://arxiv.org/abs/2107.09210}{{\ttfamily 2107.09210}}].

\bibitem{BESIII:2013ris}
{\scshape BESIII} collaboration, \emph{{Observation of a Charged Charmoniumlike
  Structure in $e^+e^-$ \textrightarrow{} $\pi^+\pi^-$ J/\ensuremath{\psi} at
  $\sqrt{s}$ =4.26 GeV}},
  \href{https://doi.org/10.1103/PhysRevLett.110.252001}{\emph{Phys. Rev. Lett.}
  {\bfseries 110} (2013) 252001}
  [\href{https://arxiv.org/abs/1303.5949}{{\ttfamily 1303.5949}}].

\bibitem{BESIII:2013ouc}
{\scshape BESIII} collaboration, \emph{{Observation of a Charged Charmoniumlike
  Structure $Z_c$(4020) and Search for the $Z_c$(3900) in $e^+e^- \to
  \pi^+\pi^-h_c$}},
  \href{https://doi.org/10.1103/PhysRevLett.111.242001}{\emph{Phys. Rev. Lett.}
  {\bfseries 111} (2013) 242001}
  [\href{https://arxiv.org/abs/1309.1896}{{\ttfamily 1309.1896}}].

\bibitem{BESIII:2013qmu}
{\scshape BESIII} collaboration, \emph{{Observation of a charged
  $(D\bar{D}^{*})^\pm$ mass peak in $e^{+}e^{-} \to \pi D\bar{D}^{*}$ at
  $\sqrt{s} =$ 4.26 GeV}},
  \href{https://doi.org/10.1103/PhysRevLett.112.022001}{\emph{Phys. Rev. Lett.}
  {\bfseries 112} (2014) 022001}
  [\href{https://arxiv.org/abs/1310.1163}{{\ttfamily 1310.1163}}].

\bibitem{BESIII:2013mhi}
{\scshape BESIII} collaboration, \emph{{Observation of a charged charmoniumlike
  structure in $e^+e^- \to (D^{*} \bar{D}^{*})^{\pm} \pi^\mp$ at
  $\sqrt{s}=4.26$GeV}},
  \href{https://doi.org/10.1103/PhysRevLett.112.132001}{\emph{Phys. Rev. Lett.}
  {\bfseries 112} (2014) 132001}
  [\href{https://arxiv.org/abs/1308.2760}{{\ttfamily 1308.2760}}].

\bibitem{BESIII:2015cld}
{\scshape BESIII} collaboration, \emph{{Observation of $Z_c(3900)^{0}$ in
  $e^+e^-\to\pi^0\pi^0 J/\psi$}},
  \href{https://doi.org/10.1103/PhysRevLett.115.112003}{\emph{Phys. Rev. Lett.}
  {\bfseries 115} (2015) 112003}
  [\href{https://arxiv.org/abs/1506.06018}{{\ttfamily 1506.06018}}].

\bibitem{BESIII:2014gnk}
{\scshape BESIII} collaboration, \emph{{Observation of $e^+e^- \to
  \pi^0\pi^0h_c$ and a Neutral Charmoniumlike Structure $Z_c(4020)^0$}},
  \href{https://doi.org/10.1103/PhysRevLett.113.212002}{\emph{Phys. Rev. Lett.}
  {\bfseries 113} (2014) 212002}
  [\href{https://arxiv.org/abs/1409.6577}{{\ttfamily 1409.6577}}].

\bibitem{BESIII:2017bua}
{\scshape BESIII} collaboration, \emph{{Determination of the Spin and Parity of
  the $Z_c(3900)$}},
  \href{https://doi.org/10.1103/PhysRevLett.119.072001}{\emph{Phys. Rev. Lett.}
  {\bfseries 119} (2017) 072001}
  [\href{https://arxiv.org/abs/1706.04100}{{\ttfamily 1706.04100}}].

\bibitem{BESIII:2020qkh}
{\scshape BESIII} collaboration, \emph{{Observation of a Near-Threshold
  Structure in the $K^+$ Recoil-Mass Spectra in $e^+e^- \rightarrow
  K^+(D_s^-D^{*0}+D_s^{*-}D^0$)}},
  \href{https://doi.org/10.1103/PhysRevLett.126.102001}{\emph{Phys. Rev. Lett.}
  {\bfseries 126} (2021) 102001}
  [\href{https://arxiv.org/abs/2011.07855}{{\ttfamily 2011.07855}}].

\bibitem{BESIII:2013fnz}
{\scshape BESIII} collaboration, \emph{{Observation of $e^+e^- \to \gamma
  X$(3872) at BESIII}},
  \href{https://doi.org/10.1103/PhysRevLett.112.092001}{\emph{Phys. Rev. Lett.}
  {\bfseries 112} (2014) 092001}
  [\href{https://arxiv.org/abs/1310.4101}{{\ttfamily 1310.4101}}].

\bibitem{BESIII:2020nbj}
{\scshape BESIII} collaboration, \emph{{Study of Open-Charm Decays and
  Radiative Transitions of the X (3872)}},
  \href{https://doi.org/10.1103/PhysRevLett.124.242001}{\emph{Phys. Rev. Lett.}
  {\bfseries 124} (2020) 242001}
  [\href{https://arxiv.org/abs/2001.01156}{{\ttfamily 2001.01156}}].

\bibitem{BESIII:2019esk}
{\scshape BESIII} collaboration, \emph{{Observation of the decay $X(3872) \to
  \pi^0 \chi_{c1}(1P)$}},
  \href{https://doi.org/10.1103/PhysRevLett.122.202001}{\emph{Phys. Rev. Lett.}
  {\bfseries 122} (2019) 202001}
  [\href{https://arxiv.org/abs/1901.03992}{{\ttfamily 1901.03992}}].

\bibitem{BESIII:2019qvy}
{\scshape BESIII} collaboration, \emph{{Study of $e^+e^- \to \gamma \omega
  J/\psi$ and Observation of $X(3872) \to \omega J/\psi$}},
  \href{https://doi.org/10.1103/PhysRevLett.122.232002}{\emph{Phys. Rev. Lett.}
  {\bfseries 122} (2019) 232002}
  [\href{https://arxiv.org/abs/1903.04695}{{\ttfamily 1903.04695}}].

\bibitem{Zhu:2021vtd}
K.~Zhu, \emph{{Triangle relations for XY Z states}},
  \href{https://doi.org/10.1142/S0217751X21501268}{\emph{Int. J. Mod. Phys. A}
  {\bfseries 36} (2021) 2150126}
  [\href{https://arxiv.org/abs/2101.10622}{{\ttfamily 2101.10622}}].

\bibitem{LHCb:2015yax}
{\scshape LHCb} collaboration, \emph{{Observation of $J/\psi p$ Resonances
  Consistent with Pentaquark States in $\Lambda_b^0 \to J/\psi K^- p$ Decays}},
  \href{https://doi.org/10.1103/PhysRevLett.115.072001}{\emph{Phys. Rev. Lett.}
  {\bfseries 115} (2015) 072001}
  [\href{https://arxiv.org/abs/1507.03414}{{\ttfamily 1507.03414}}].

\bibitem{LHCb:2019kea}
{\scshape LHCb} collaboration, \emph{{Observation of a narrow pentaquark state,
  $P_c(4312)^+$, and of two-peak structure of the $P_c(4450)^+$}},
  \href{https://doi.org/10.1103/PhysRevLett.122.222001}{\emph{Phys. Rev. Lett.}
  {\bfseries 122} (2019) 222001}
  [\href{https://arxiv.org/abs/1904.03947}{{\ttfamily 1904.03947}}].

\bibitem{Shen:2015eua}
C.-P.~Shen and C.-Z.~Yuan, \emph{{Observation of pentaquark states and
  perspectives of further studies}},
  \href{https://doi.org/10.1007/s11434-015-0881-1}{\emph{Science Bulletin}
  {\bfseries 60} (2015) 1549}
  [\href{https://arxiv.org/abs/1508.06047}{{\ttfamily 1508.06047}}].

\bibitem{pdg2020}
{\scshape Particle Data Group} collaboration, \emph{{Review of Particle
  Physics}}, \href{https://doi.org/10.1093/ptep/ptaa104}{\emph{PTEP} {\bfseries
  2020} (2020) 083C01}.

\bibitem{prd98_074512}
A.~Bazavov et~al., \emph{{$B$- and $D$-meson leptonic decay constants from
  four-flavor lattice QCD}},
  \href{https://doi.org/10.1103/PhysRevD.98.074512}{\emph{Phys. Rev. D}
  {\bfseries 98} (2018) 074512}
  [\href{https://arxiv.org/abs/1712.09262}{{\ttfamily 1712.09262}}].

\bibitem{prd91_054507}
N.~Carrasco et~al., \emph{{Leptonic decay constants $f_{K},f_{D},$ and
  $f_{{D}_{s}}$ with $N_{f} = 2+1+1$ twisted-mass lattice QCD}},
  \href{https://doi.org/10.1103/PhysRevD.91.054507}{\emph{Phys. Rev. D}
  {\bfseries 91} (2015) 054507}
  [\href{https://arxiv.org/abs/1411.7908}{{\ttfamily 1411.7908}}].

\bibitem{bes3_muv}
{\scshape BESIII} collaboration, \emph{{Precision measurements of $B(D^+
  \rightarrow \mu^+ \nu_{\mu})$, the pseudoscalar decay constant $f_{D^+}$, and
  the quark mixing matrix element $|V_{\rm cd}|$}},
  \href{https://doi.org/10.1103/PhysRevD.89.051104}{\emph{Phys. Rev. D}
  {\bfseries 89} (2014) 051104}
  [\href{https://arxiv.org/abs/1312.0374}{{\ttfamily 1312.0374}}].

\bibitem{Ablikim:2019rpl}
{\scshape BESIII} collaboration, \emph{{Observation of the leptonic decay $D^+
  \to \tau^+ \nu_\tau$}},
  \href{https://doi.org/10.1103/PhysRevLett.123.211802}{\emph{Phys. Rev. Lett.}
  {\bfseries 123} (2019) 211802}
  [\href{https://arxiv.org/abs/1908.08877}{{\ttfamily 1908.08877}}].

\bibitem{bes3_Ds_muv}
{\scshape BESIII} collaboration, \emph{{Determination of the pseudoscalar decay
  constant $f_{D_s^+}$ via $D_s^+\to\mu^+\nu_\mu$}},
  \href{https://doi.org/10.1103/PhysRevLett.122.071802}{\emph{Phys. Rev. Lett.}
  {\bfseries 122} (2019) 071802}
  [\href{https://arxiv.org/abs/1811.10890}{{\ttfamily 1811.10890}}].

\bibitem{bes3_Ds_tauv1}
{\scshape BESIII} collaboration, \emph{{Measurement of the branching fraction
  of leptonic decay $D_s^+\to\tau^+\nu_\tau$ via $\tau^+\to\pi^+\pi^0\bar
  \nu_\tau$}}, \href{https://doi.org/10.1103/PhysRevD.104.032001}{\emph{Phys.
  Rev. D} {\bfseries 104} (2021) 032001}
  [\href{https://arxiv.org/abs/2105.07178}{{\ttfamily 2105.07178}}].

\bibitem{bes3_Ds_tauv2}
{\scshape BESIII} collaboration, \emph{{Measurement of the absolute branching
  fractions for purely leptonic $D_s^+$ decays}},
  \href{https://doi.org/10.1103/PhysRevD.104.052009}{\emph{Phys. Rev. D}
  {\bfseries 104} (2021) 052009}
  [\href{https://arxiv.org/abs/2102.11734}{{\ttfamily 2102.11734}}].

\bibitem{bes3_Ds_tauv3}
{\scshape BESIII} collaboration, \emph{{Measurement of the Absolute Branching
  Fraction of $D_s^+ \to \tau^+ \nu_{\tau}$ via $\tau^+ \to e^+ \nu_e
  \bar{\nu}_{\tau}$}},
  \href{https://doi.org/10.1103/PhysRevLett.127.171801}{\emph{Phys. Rev. Lett.}
  {\bfseries 127} (2021) 171801}
  [\href{https://arxiv.org/abs/2106.02218}{{\ttfamily 2106.02218}}].

\bibitem{Ivanov:2019nqd}
M.A.~Ivanov, J.G.~K\"orner, J.N.~Pandya, P.~Santorelli, N.R.~Soni and
  C.-T.~Tran, \emph{{Exclusive semileptonic decays of D and D$_{s}$ mesons in
  the covariant confining quark model}},
  \href{https://doi.org/10.1007/s11467-019-0908-1}{\emph{Front. Phys.
  (Beijing)} {\bfseries 14} (2019) 64401}
  [\href{https://arxiv.org/abs/1904.07740}{{\ttfamily 1904.07740}}].

\bibitem{bes3_D0_kpiev}
{\scshape BESIII} collaboration, \emph{{Study of Dynamics of $D^0 \to K^- e^+
  \nu_{e}$ and $D^0\to\pi^- e^+ \nu_{e}$ Decays}},
  \href{https://doi.org/10.1103/PhysRevD.92.072012}{\emph{Phys. Rev. D}
  {\bfseries 92} (2015) 072012}
  [\href{https://arxiv.org/abs/1508.07560}{{\ttfamily 1508.07560}}].

\bibitem{bes3_Dp_k0pi0ev}
{\scshape BESIII} collaboration, \emph{{Analysis of $D^+\to\bar K^0e^+\nu_e$
  and $D^+\to\pi^0e^+\nu_e$ semileptonic decays}},
  \href{https://doi.org/10.1103/PhysRevD.96.012002}{\emph{Phys. Rev. D}
  {\bfseries 96} (2017) 012002}
  [\href{https://arxiv.org/abs/1703.09084}{{\ttfamily 1703.09084}}].

\bibitem{bes3_Dp_KLev}
{\scshape BESIII} collaboration, \emph{{Study of decay dynamics and $CP$
  asymmetry in $D^+ \to K^0_L e^+ \nu_e$ decay}},
  \href{https://doi.org/10.1103/PhysRevD.92.112008}{\emph{Phys. Rev. D}
  {\bfseries 92} (2015) 112008}
  [\href{https://arxiv.org/abs/1510.00308}{{\ttfamily 1510.00308}}].

\bibitem{bes3_D_kev}
{\scshape BESIII} collaboration, \emph{{Determination of the absolute branching
  fractions of $D^0\to K^-e^+\nu_e$ and $D^+\to \bar K^0 e^+\nu_e$}},
  \href{https://doi.org/10.1103/PhysRevD.104.052008}{\emph{Phys. Rev. D}
  {\bfseries 104} (2021) 052008}
  [\href{https://arxiv.org/abs/2104.08081}{{\ttfamily 2104.08081}}].

\bibitem{bes3_Dp_KSev_2pi0}
{\scshape BESIII} collaboration, \emph{{Measurement of the absolute branching
  fraction of $D^{+}\rightarrow\bar K^0 e^{+}\nu_{e}$ via $\bar
  K^0\to\pi^0\pi^0$}},
  \href{https://doi.org/10.1088/1674-1137/40/11/113001}{\emph{Chin. Phys. C}
  {\bfseries 40} (2016) 113001}
  [\href{https://arxiv.org/abs/1605.00208}{{\ttfamily 1605.00208}}].

\bibitem{bes3_D0_kmuv}
{\scshape BESIII} collaboration, \emph{{Study of the $D^0\to K^-\mu^+\nu_\mu$
  dynamics and test of lepton flavor universality with $D^0\to
  K^-\ell^+\nu_\ell$ decays}},
  \href{https://doi.org/10.1103/PhysRevLett.122.011804}{\emph{Phys. Rev. Lett.}
  {\bfseries 122} (2019) 011804}
  [\href{https://arxiv.org/abs/1810.03127}{{\ttfamily 1810.03127}}].

\bibitem{bes3_Dp_kmuv}
{\scshape BESIII} collaboration, \emph{{Improved measurement of the absolute
  branching fraction of $D^{+}\rightarrow \bar{K}^0 \mu ^{+}\nu _{\mu }$}},
  \href{https://doi.org/10.1140/epjc/s10052-016-4198-2}{\emph{Eur. Phys. J. C}
  {\bfseries 76} (2016) 369}
  [\href{https://arxiv.org/abs/1605.00068}{{\ttfamily 1605.00068}}].

\bibitem{bes3_Dst_Kev}
{\scshape BESIII} collaboration, \emph{{First Measurement of the Form Factors
  in $D^+_{s}\rightarrow K^0 e^+\nu_e$ and $D^+_{s}\rightarrow K^{*0} e^+\nu_e$
  Decays}}, \href{https://doi.org/10.1103/PhysRevLett.122.061801}{\emph{Phys.
  Rev. Lett.} {\bfseries 122} (2019) 061801}
  [\href{https://arxiv.org/abs/1811.02911}{{\ttfamily 1811.02911}}].

\bibitem{bes3_D_pimuv}
{\scshape BESIII} collaboration, \emph{{Measurement of the branching fraction
  for the semi-leptonic decay $D^{0(+)}\to \pi^{-(0)}\mu^+\nu_\mu$ and test of
  lepton universality}},
  \href{https://doi.org/10.1103/PhysRevLett.121.171803}{\emph{Phys. Rev. Lett.}
  {\bfseries 121} (2018) 171803}
  [\href{https://arxiv.org/abs/1802.05492}{{\ttfamily 1802.05492}}].

\bibitem{bes3_Dp_etaev}
{\scshape BESIII} collaboration, \emph{{Study of the decays
  $D^+\rightarrow\eta^{(\prime)} e^+\nu_{e}$}},
  \href{https://doi.org/10.1103/PhysRevD.97.092009}{\emph{Phys. Rev. D}
  {\bfseries 97} (2018) 092009}
  [\href{https://arxiv.org/abs/1803.05570}{{\ttfamily 1803.05570}}].

\bibitem{bes3_Dp_etamuv}
{\scshape BESIII} collaboration, \emph{{First Observation of $D^+ \rightarrow
  \eta\mu^+\nu_\mu$ and Measurement of Its Decay Dynamics}},
  \href{https://doi.org/10.1103/PhysRevLett.124.231801}{\emph{Phys. Rev. Lett.}
  {\bfseries 124} (2020) 231801}
  [\href{https://arxiv.org/abs/2003.12220}{{\ttfamily 2003.12220}}].

\bibitem{bes3_Dst_etaenu}
{\scshape BESIII} collaboration, \emph{{Measurement of the Dynamics of the
  Decays $D_s^+ \rightarrow \eta^{(\prime)} e^+ \nu_e$}},
  \href{https://doi.org/10.1103/PhysRevLett.122.121801}{\emph{Phys. Rev. Lett.}
  {\bfseries 122} (2019) 121801}
  [\href{https://arxiv.org/abs/1901.02133}{{\ttfamily 1901.02133}}].

\bibitem{bes3_Ds_etaev_4009}
{\scshape BESIII} collaboration, \emph{{Measurements of the absolute branching
  fractions for $D_{s}^{+}\rightarrow\eta e^{+}\nu_{e}$ and
  $D_{s}^{+}\rightarrow\eta^{\prime} e^{+}\nu_{e}$}},
  \href{https://doi.org/10.1103/PhysRevD.94.112003}{\emph{Phys. Rev. D}
  {\bfseries 94} (2016) 112003}
  [\href{https://arxiv.org/abs/1608.06484}{{\ttfamily 1608.06484}}].

\bibitem{bes3_Ds_etamuv_4009}
{\scshape BESIII} collaboration, \emph{{Measurements of the branching fractions
  for the semi-leptonic decays $D^+_s\to\phi e^{+}\nu_{e}$, $\phi
  \mu^{+}\nu_{\mu}$, $\eta \mu^{+}\nu_{\mu}$ and $\eta'\mu^{+}\nu_{\mu}$}},
  \href{https://doi.org/10.1103/PhysRevD.97.012006}{\emph{Phys. Rev. D}
  {\bfseries 97} (2018) 012006}
  [\href{https://arxiv.org/abs/1709.03680}{{\ttfamily 1709.03680}}].

\bibitem{bes3_Dp_kpiev}
{\scshape BESIII} collaboration, \emph{{Study of $D^{+} \to K^{-} \pi^+ e^+
  \nu_e$}}, \href{https://doi.org/10.1103/PhysRevD.94.032001}{\emph{Phys. Rev.
  D} {\bfseries 94} (2016) 032001}
  [\href{https://arxiv.org/abs/1512.08627}{{\ttfamily 1512.08627}}].

\bibitem{bes3_D0_kspiev}
{\scshape BESIII} collaboration, \emph{{Study of the decay $D^0\rightarrow
  \bar{K}^0\pi^-e^+\nu_e$}},
  \href{https://doi.org/10.1103/PhysRevD.99.011103}{\emph{Phys. Rev. D}
  {\bfseries 99} (2019) 011103}
  [\href{https://arxiv.org/abs/1811.11349}{{\ttfamily 1811.11349}}].

\bibitem{bes3_D0_pipiev}
{\scshape BESIII} collaboration, \emph{{Observation of $D^+ \to f_0(500)
  e^+\nu_e$ and Improved Measurements of $D \to\rho e^+\nu_e$}},
  \href{https://doi.org/10.1103/PhysRevLett.122.062001}{\emph{Phys. Rev. Lett.}
  {\bfseries 122} (2019) 062001}
  [\href{https://arxiv.org/abs/1809.06496}{{\ttfamily 1809.06496}}].

\bibitem{bes3_D0_rhomuv}
{\scshape BESIII} collaboration, \emph{{Observation of the decay $D^0\to
  \rho^-\mu^+\nu_\mu$}},
  \href{https://doi.org/10.1103/PhysRevD.104.L091103}{\emph{Phys. Rev. D}
  {\bfseries 104} (2021) L091103}
  [\href{https://arxiv.org/abs/2106.02292}{{\ttfamily 2106.02292}}].

\bibitem{bes3_Dp_omegaev}
{\scshape BESIII} collaboration, \emph{{Measurement of the form factors in the
  decay $D^+ \to \omega e^+ \nu_{e}$ and search for the decay $D^+ \to \phi e^+
  \nu_{e}$}}, \href{https://doi.org/10.1103/PhysRevD.92.071101}{\emph{Phys.
  Rev. D} {\bfseries 92} (2015) 071101}
  [\href{https://arxiv.org/abs/1508.00151}{{\ttfamily 1508.00151}}].

\bibitem{bes3_Dp_omegamuv}
{\scshape BESIII} collaboration, \emph{{Observation of the semimuonic decay
  $D^+\to \omega\mu^+\nu_\mu$}},
  \href{https://doi.org/10.1103/PhysRevD.101.072005}{\emph{Phys. Rev. D}
  {\bfseries 101} (2020) 072005}
  [\href{https://arxiv.org/abs/2002.10578}{{\ttfamily 2002.10578}}].

\bibitem{bes3_Ds_f0enu}
{\scshape BESIII} collaboration, \emph{{Study of light scalar mesons through
  $D^+_s \to \pi^0\pi^0e^+ \nu_e$ and $K^0_S K^0_S e^+ \nu_e$ decays}},
  \href{https://doi.org/10.1103/PhysRevD.105.L031101}{\emph{Phys. Rev. D}
  {\bfseries 105} (2022) L031101}.

\bibitem{bes3_D_a0ev}
{\scshape BESIII} collaboration, \emph{{Observation of the Semileptonic Decay
  $D^0 \to a_0(980)^- e^+ \nu_e$ and Evidence for $D^+ \to a_0(980)^0 e^+
  \nu_e$}}, \href{https://doi.org/10.1103/PhysRevLett.121.081802}{\emph{Phys.
  Rev. Lett.} {\bfseries 121} (2018) 081802}
  [\href{https://arxiv.org/abs/1803.02166}{{\ttfamily 1803.02166}}].

\bibitem{bes3_Ds_a0enu}
{\scshape BESIII} collaboration, \emph{{Search for the decay $D_s^+\to
  a_0(980)^0e^+\nu_e$}},
  \href{https://doi.org/10.1103/PhysRevD.103.092004}{\emph{Phys. Rev. D}
  {\bfseries 103} (2021) 092004}
  [\href{https://arxiv.org/abs/2103.11855}{{\ttfamily 2103.11855}}].

\bibitem{bes3_Dp_K1enu}
{\scshape BESIII} collaboration, \emph{{Observation of the Semileptonic $D^+$
  Decay into the $\bar K_1(1270)^0$ Axial-Vector Meson}},
  \href{https://doi.org/10.1103/PhysRevLett.123.231801}{\emph{Phys. Rev. Lett.}
  {\bfseries 123} (2019) 231801}
  [\href{https://arxiv.org/abs/1907.11370}{{\ttfamily 1907.11370}}].

\bibitem{bes3_D0_K1enu}
{\scshape BESIII} collaboration, \emph{{Observation of $D^0\to K_1(1270)^-
  e^+\nu_e$}},
  \href{https://doi.org/10.1103/PhysRevLett.127.131801}{\emph{Phys. Rev. Lett.}
  {\bfseries 127} (2021) 131801}
  [\href{https://arxiv.org/abs/2102.10850}{{\ttfamily 2102.10850}}].

\bibitem{bes3_D_b1enu}
{\scshape BESIII} collaboration, \emph{{Search for the semileptonic decay
  $D^{0(+)}\to b_1(1235)^{-(0)} e^+\nu_e$}},
  \href{https://doi.org/10.1103/PhysRevD.102.112005}{\emph{Phys. Rev. D}
  {\bfseries 102} (2020) 112005}
  [\href{https://arxiv.org/abs/2008.05754}{{\ttfamily 2008.05754}}].

\bibitem{lqcd_fk}
H.~Na, C.T.H.~Davies, E.~Follana, G.P.~Lepage and J.~Shigemitsu, \emph{{The $D
  \rightarrow K, l \nu$ Semileptonic Decay Scalar Form Factor and $|V_{cs}|$
  from Lattice QCD}},
  \href{https://doi.org/10.1103/PhysRevD.82.114506}{\emph{Phys. Rev. D}
  {\bfseries 82} (2010) 114506}
  [\href{https://arxiv.org/abs/1008.4562}{{\ttfamily 1008.4562}}].

\bibitem{lqcd_fpi}
H.~Na, C.T.H.~Davies, E.~Follana, J.~Koponen, G.P.~Lepage and J.~Shigemitsu,
  \emph{{$D \rightarrow \pi, l \nu$ Semileptonic Decays, $|V_{cd}|$ and
  2$^{nd}$ Row Unitarity from Lattice QCD}},
  \href{https://doi.org/10.1103/PhysRevD.84.114505}{\emph{Phys. Rev. D}
  {\bfseries 84} (2011) 114505}
  [\href{https://arxiv.org/abs/1109.1501}{{\ttfamily 1109.1501}}].

\bibitem{lqcd_ETM}
{\scshape ETM} collaboration, \emph{{Scalar and vector form factors of $D \to
  \pi(K) \ell \nu$ decays with $N_f=2+1+1$ twisted fermions}},
  \href{https://doi.org/10.1103/PhysRevD.96.054514}{\emph{Phys. Rev. D}
  {\bfseries 96} (2017) 054514}
  [\href{https://arxiv.org/abs/1706.03017}{{\ttfamily 1706.03017}}].

\bibitem{lqcd_MILC}
{\scshape Fermilab Lattice, MILC} collaboration, \emph{{$D$ meson Semileptonic
  Decay Form Factors at $q^2 = 0$}},
  \href{https://doi.org/10.22323/1.334.0269}{\emph{PoS} {\bfseries LATTICE2018}
  (2019) 269} [\href{https://arxiv.org/abs/1901.08989}{{\ttfamily
  1901.08989}}].

\bibitem{BROD}
J.~Brod and J.~Zupan, \emph{{The ultimate theoretical error on $\gamma$ from $B
  \to DK$ decays}}, \href{https://doi.org/10.1007/JHEP01(2014)051}{\emph{JHEP}
  {\bfseries 01} (2014) 051} [\href{https://arxiv.org/abs/1308.5663}{{\ttfamily
  1308.5663}}].

\bibitem{GGSZ}
A.~Giri, Y.~Grossman, A.~Soffer and J.~Zupan, \emph{{Determining gamma using
  $B^\pm \to DK^\pm$ with multibody D decays}},
  \href{https://doi.org/10.1103/PhysRevD.68.054018}{\emph{Phys. Rev. D}
  {\bfseries 68} (2003) 054018}
  [\href{https://arxiv.org/abs/hep-ph/0303187}{{\ttfamily hep-ph/0303187}}].

\bibitem{BESIIIcisi}
{\scshape BESIII} collaboration, \emph{{Model-independent determination of the
  relative strong-phase difference between $D^0$ and $\bar{D}^0\rightarrow
  K^0_{S,L}\pi^+\pi^-$ and its impact on the measurement of the CKM angle
  $\gamma/\phi_3$}},
  \href{https://doi.org/10.1103/PhysRevD.101.112002}{\emph{Phys. Rev. D}
  {\bfseries 101} (2020) 112002}
  [\href{https://arxiv.org/abs/2003.00091}{{\ttfamily 2003.00091}}].

\bibitem{lhcb:gamma}
{\scshape LHCb} collaboration, \emph{{Measurement of the CKM angle $\gamma$ in
  $B^\pm\to D K^\pm$ and $B^\pm \to D \pi^\pm$ decays with $D \to K_\mathrm S^0
  h^+ h^-$}}, \href{https://doi.org/10.1007/JHEP02(2021)169}{\emph{JHEP}
  {\bfseries 02} (2021) 169}
  [\href{https://arxiv.org/abs/2010.08483}{{\ttfamily 2010.08483}}].

\bibitem{belle:gamma}
{\scshape Belle, Belle-II} collaboration, \emph{{Combined analysis of Belle and
  Belle II data to determine the CKM angle $ \phi_{3} $ using $B^+ \to
  D(K_{S}^0 h^- h^+) h^+$ decays}},
  \href{https://doi.org/10.1007/JHEP02(2022)063}{\emph{JHEP} {\bfseries 02}
  (2022) 063} [\href{https://arxiv.org/abs/2110.12125}{{\ttfamily
  2110.12125}}].

\bibitem{binflip}
{\scshape LHCb} collaboration, \emph{{Observation of the Mass Difference
  Between Neutral Charm-Meson Eigenstates}},
  \href{https://doi.org/10.1103/PhysRevLett.127.111801}{\emph{Phys. Rev. Lett.}
  {\bfseries 127} (2021) 111801}
  [\href{https://arxiv.org/abs/2106.03744}{{\ttfamily 2106.03744}}].

\bibitem{Muong-2:2021ojo}
{\scshape Muon g-2} collaboration, \emph{{Measurement of the Positive Muon
  Anomalous Magnetic Moment to 0.46 ppm}},
  \href{https://doi.org/10.1103/PhysRevLett.126.141801}{\emph{Phys. Rev. Lett.}
  {\bfseries 126} (2021) 141801}
  [\href{https://arxiv.org/abs/2104.03281}{{\ttfamily 2104.03281}}].

\bibitem{Aoyama:2020ynm}
T.~Aoyama et~al., \emph{{The anomalous magnetic moment of the muon in the
  Standard Model}},
  \href{https://doi.org/10.1016/j.physrep.2020.07.006}{\emph{Phys. Rept.}
  {\bfseries 887} (2020) 1} [\href{https://arxiv.org/abs/2006.04822}{{\ttfamily
  2006.04822}}].

\bibitem{BESIII:2015equ}
{\scshape BESIII} collaboration, \emph{{Measurement of the $e^+ e^- \to \pi^+
  \pi^-$ cross section between 600 and 900 MeV using initial state radiation}},
  \href{https://doi.org/10.1016/j.physletb.2015.11.043}{\emph{Phys. Lett. B}
  {\bfseries 753} (2016) 629}
  [\href{https://arxiv.org/abs/1507.08188}{{\ttfamily 1507.08188}}].

\bibitem{BESIII:2019gjz}
{\scshape BESIII} collaboration, \emph{{Measurement of the
  $e^+e^-\to\pi^+\pi^-\pi^0$ Cross Section from 0.7 GeV to 3.0 GeV via
  Initial-State Radiation}},
  \href{https://arxiv.org/abs/1912.11208}{{\ttfamily 1912.11208}}.

\bibitem{Redmer:2018nxj}
{\scshape BESIII} collaboration, \emph{{Measurement of Hadronic Cross Sections
  at BESIII}},  in \emph{{13th Conference on the Intersections of Particle and
  Nuclear Physics}}, 10, 2018
  [\href{https://arxiv.org/abs/1810.00643}{{\ttfamily 1810.00643}}].

\bibitem{BESIII:2021wib}
{\scshape BESIII} collaboration, \emph{{Measurement of the Cross Section for
  $e^{+}e^{-}\to$Hadrons at Energies from 2.2324 to 3.6710~GeV}},
  \href{https://doi.org/10.1103/PhysRevLett.128.062004}{\emph{Phys. Rev. Lett.}
  {\bfseries 128} (2022) 062004}
  [\href{https://arxiv.org/abs/2112.11728}{{\ttfamily 2112.11728}}].

\bibitem{Ping:2016pms}
R.-G.~Ping et~al., \emph{{Tuning and validation of hadronic event generator for
  $R$ value measurements in the tau-charm region}},
  \href{https://doi.org/10.1088/1674-1137/40/11/113002}{\emph{Chin. Phys. C}
  {\bfseries 40} (2016) 113002}
  [\href{https://arxiv.org/abs/1605.09208}{{\ttfamily 1605.09208}}].

\bibitem{BESIII:2020gnc}
{\scshape BESIII} collaboration, \emph{{Observation of a structure in
  $e^{+}e^{-} \to \phi \eta^{\prime}$ at $\sqrt{s}$ from 2.05 to 3.08 GeV}},
  \href{https://doi.org/10.1103/PhysRevD.102.012008}{\emph{Phys. Rev. D}
  {\bfseries 102} (2020) 012008}
  [\href{https://arxiv.org/abs/2003.13064}{{\ttfamily 2003.13064}}].

\bibitem{BESIII:2021bjn}
{\scshape BESIII} collaboration, \emph{{Study of the process
  $e^{+}e^{-}\rightarrow\phi\eta$ at center-of-mass energies between 2.00 and
  3.08 GeV}}, \href{https://doi.org/10.1103/PhysRevD.104.032007}{\emph{Phys.
  Rev. D} {\bfseries 104} (2021) 032007}
  [\href{https://arxiv.org/abs/2104.05549}{{\ttfamily 2104.05549}}].

\bibitem{BESIII:2018ldc}
{\scshape BESIII} collaboration, \emph{{Measurement of $e^{+} e^{-} \rightarrow
  K^{+} K^{-}$ cross section at $\sqrt{s} = 2.00 - 3.08$ GeV}},
  \href{https://doi.org/10.1103/PhysRevD.99.032001}{\emph{Phys. Rev. D}
  {\bfseries 99} (2019) 032001}
  [\href{https://arxiv.org/abs/1811.08742}{{\ttfamily 1811.08742}}].

\bibitem{BESIII:2021yam}
{\scshape BESIII} collaboration, \emph{{Cross section measurement of
  $e^{+}e^{-} \to K_{S}^{0}K_{L}^{0}$ at $\sqrt{s}=2.00-3.08~{GeV}$}},
  \href{https://doi.org/10.1103/PhysRevD.104.092014}{\emph{Phys. Rev. D}
  {\bfseries 104} (2021) 092014}
  [\href{https://arxiv.org/abs/2105.13597}{{\ttfamily 2105.13597}}].

\bibitem{BESIII:2020vtu}
{\scshape BESIII} collaboration, \emph{{Observation of a Resonant Structure in
  $e^{+}e^{-} \to K^{+}K^{-}\pi^{0}\pi^{0}$}},
  \href{https://doi.org/10.1103/PhysRevLett.124.112001}{\emph{Phys. Rev. Lett.}
  {\bfseries 124} (2020) 112001}
  [\href{https://arxiv.org/abs/2001.04131}{{\ttfamily 2001.04131}}].

\bibitem{BESIII:2022wxz}
{\scshape BESIII} collaboration, \emph{{Measurement of $e^{+}e^{-} \to
  K^{+}K^{-}\pi^{0}$ cross section and observation of a resonant structure}},
  \href{https://arxiv.org/abs/2202.06447}{{\ttfamily 2202.06447}}.

\bibitem{BESIII:2020xmw}
{\scshape BESIII} collaboration, \emph{{Observation of a resonant structure in
  $e^{+}e^{-} \to \omega\eta$ and another in $e^{+}e^{-} \to \omega\pi^{0}$ at
  center-of-mass energies between 2.00 and 3.08 GeV}},
  \href{https://doi.org/10.1016/j.physletb.2020.136059}{\emph{Phys. Lett. B}
  {\bfseries 813} (2021) 136059}
  [\href{https://arxiv.org/abs/2009.08099}{{\ttfamily 2009.08099}}].

\bibitem{BESIII:2020kpr}
{\scshape BESIII} collaboration, \emph{{Measurement of the Born cross sections
  for $e^+e^- \to \eta^\prime \pi^{+}\pi^{-}$ at center-of-mass energies
  between $2.00$ and $3.08$\textasciitilde{}GeV}},
  \href{https://doi.org/10.1103/PhysRevD.103.072007}{\emph{Phys. Rev. D}
  {\bfseries 103} (2021) 072007}
  [\href{https://arxiv.org/abs/2012.07360}{{\ttfamily 2012.07360}}].

\bibitem{BESIII:2021uni}
{\scshape BESIII} collaboration, \emph{{Measurement of the
  $e^{+}e^{-}\rightarrow\omega\pi^{0}\pi^{0}$ cross section at center-of-mass
  energies from 2.0 to 3.08~GeV}},
  \href{https://doi.org/10.1103/PhysRevD.105.032005}{\emph{Phys. Rev. D}
  {\bfseries 105} (2022) 032005}
  [\href{https://arxiv.org/abs/2112.15076}{{\ttfamily 2112.15076}}].

\bibitem{Punjabi:2015bba}
V.~Punjabi, C.F.~Perdrisat, M.K.~Jones, E.J.~Brash and C.E.~Carlson, \emph{{The
  Structure of the Nucleon: Elastic Electromagnetic Form Factors}},
  \href{https://doi.org/10.1140/epja/i2015-15079-x}{\emph{Eur. Phys. J. A}
  {\bfseries 51} (2015) 79} [\href{https://arxiv.org/abs/1503.01452}{{\ttfamily
  1503.01452}}].

\bibitem{Puckett:2011xg}
A.J.R.~Puckett et~al., \emph{{Final Analysis of Proton Form Factor Ratio Data
  at $\mathbf{Q^2 = 4.0}$, 4.8 and 5.6 GeV$\mathbf{^2}$}},
  \href{https://doi.org/10.1103/PhysRevC.85.045203}{\emph{Phys. Rev. C}
  {\bfseries 85} (2012) 045203}
  [\href{https://arxiv.org/abs/1102.5737}{{\ttfamily 1102.5737}}].

\bibitem{BESIII:2015axk}
{\scshape BESIII} collaboration, \emph{{Measurement of the proton form factor
  by studying $e^{+} e^{-}\rightarrow p\bar{p}$}},
  \href{https://doi.org/10.1103/PhysRevD.91.112004}{\emph{Phys. Rev. D}
  {\bfseries 91} (2015) 112004}
  [\href{https://arxiv.org/abs/1504.02680}{{\ttfamily 1504.02680}}].

\bibitem{BESIII:2019hdp}
{\scshape BESIII} collaboration, \emph{{Measurement of proton electromagnetic
  form factors in $e^+e^- \to p\bar{p}$ in the energy region 2.00 - 3.08 GeV}},
  \href{https://doi.org/10.1103/PhysRevLett.124.042001}{\emph{Phys. Rev. Lett.}
  {\bfseries 124} (2020) 042001}
  [\href{https://arxiv.org/abs/1905.09001}{{\ttfamily 1905.09001}}].

\bibitem{BESIII:2019tgo}
{\scshape BESIII} collaboration, \emph{{Study of the process $e^+ e^- \to p
  \bar p$ via initial state radiation at BESIII}},
  \href{https://doi.org/10.1103/PhysRevD.99.092002}{\emph{Phys. Rev. D}
  {\bfseries 99} (2019) 092002}
  [\href{https://arxiv.org/abs/1902.00665}{{\ttfamily 1902.00665}}].

\bibitem{BESIII:2021rqk}
{\scshape BESIII} collaboration, \emph{{Measurement of proton electromagnetic
  form factors in the time-like region using initial state radiation at
  BESIII}}, \href{https://doi.org/10.1016/j.physletb.2021.136328}{\emph{Phys.
  Lett. B} {\bfseries 817} (2021) 136328}
  [\href{https://arxiv.org/abs/2102.10337}{{\ttfamily 2102.10337}}].

\bibitem{BESIII:2021tbq}
{\scshape BESIII} collaboration, \emph{{Oscillating features in the
  electromagnetic structure of the neutron}},
  \href{https://doi.org/10.1038/s41567-021-01345-6}{\emph{Nature Phys.}
  {\bfseries 17} (2021) 1200}.

\bibitem{Chernyak:1983ej}
V.L.~Chernyak and A.R.~Zhitnitsky, \emph{{Asymptotic Behavior of Exclusive
  Processes in QCD}},
  \href{https://doi.org/10.1016/0370-1573(84)90126-1}{\emph{Phys. Rept.}
  {\bfseries 112} (1984) 173}.

\bibitem{Ellis:2001xc}
J.R.~Ellis and M.~Karliner, \emph{{On electron positron annihilation into
  nucleon anti-nucleon pairs}},
  \href{https://doi.org/10.1088/1367-2630/4/1/318}{\emph{New J. Phys.}
  {\bfseries 4} (2002) 18}
  [\href{https://arxiv.org/abs/hep-ph/0108259}{{\ttfamily hep-ph/0108259}}].

\bibitem{Antonelli:1998fv}
A.~Antonelli et~al., \emph{{The first measurement of the neutron
  electromagnetic form-factors in the timelike region}},
  \href{https://doi.org/10.1016/S0550-3213(98)00083-2}{\emph{Nucl. Phys. B}
  {\bfseries 517} (1998) 3}.

\bibitem{BaBar:2013ves}
{\scshape BaBar} collaboration, \emph{{Study of $e^+e^- \to p \bar{p}$ via
  initial-state radiation at BABAR}},
  \href{https://doi.org/10.1103/PhysRevD.87.092005}{\emph{Phys. Rev. D}
  {\bfseries 87} (2013) 092005}
  [\href{https://arxiv.org/abs/1302.0055}{{\ttfamily 1302.0055}}].

\bibitem{BESIII:2017hyw}
{\scshape BESIII} collaboration, \emph{{Observation of a cross-section
  enhancement near mass threshold in
  $e^{+}e^{-}\rightarrow\Lambda\bar{\Lambda}$}},
  \href{https://doi.org/10.1103/PhysRevD.97.032013}{\emph{Phys. Rev. D}
  {\bfseries 97} (2018) 032013}
  [\href{https://arxiv.org/abs/1709.10236}{{\ttfamily 1709.10236}}].

\bibitem{BESIII:2020uqk}
{\scshape BESIII} collaboration, \emph{{Measurements of $\Sigma^+$ and
  $\Sigma^-$ time-like electromagnetic form factors for center-of-mass energies
  from 2.3864 to 3.0200 GeV}},
  \href{https://doi.org/10.1016/j.physletb.2021.136110}{\emph{Phys. Lett. B}
  {\bfseries 814} (2021) 136110}
  [\href{https://arxiv.org/abs/2009.01404}{{\ttfamily 2009.01404}}].

\bibitem{BESIII:2021rkn}
{\scshape BESIII} collaboration, \emph{{Measurement of the
  $e^{+}e^{-}\to\Sigma^{0}\bar{\Sigma}^{0}$ cross sections at center-of-mass
  energies from $2.3864$ to $3.0200$ GeV}},
  \href{https://arxiv.org/abs/2110.04510}{{\ttfamily 2110.04510}}.

\bibitem{BESIII:2020ktn}
{\scshape BESIII} collaboration, \emph{{Measurement of cross section for
  $e^+e^-\to\Xi^-\bar{\Xi}^+$ near threshold at BESIII}},
  \href{https://doi.org/10.1103/PhysRevD.103.012005}{\emph{Phys. Rev. D}
  {\bfseries 103} (2021) 012005}
  [\href{https://arxiv.org/abs/2010.08320}{{\ttfamily 2010.08320}}].

\bibitem{BESIII:2021aer}
{\scshape BESIII} collaboration, \emph{{Measurement of cross section for
  $e^{+}e^{-}\rightarrow\Xi^{0}\bar{\Xi}^{0}$ near threshold}},
  \href{https://doi.org/10.1016/j.physletb.2021.136557}{\emph{Phys. Lett. B}
  {\bfseries 820} (2021) 136557}
  [\href{https://arxiv.org/abs/2105.14657}{{\ttfamily 2105.14657}}].

\bibitem{BESIII:2017kqg}
{\scshape BESIII} collaboration, \emph{{Precision measurement of the
  $e^{+}e^{-}~\rightarrow~\Lambda_{c}^{+} \bar{\Lambda}_{c}^{-}$ cross section
  near threshold}},
  \href{https://doi.org/10.1103/PhysRevLett.120.132001}{\emph{Phys. Rev. Lett.}
  {\bfseries 120} (2018) 132001}
  [\href{https://arxiv.org/abs/1710.00150}{{\ttfamily 1710.00150}}].

\bibitem{Cabibbo:1961sz}
N.~Cabibbo and R.~Gatto, \emph{{Electron Positron Colliding Beam Experiments}},
  \href{https://doi.org/10.1103/PhysRev.124.1577}{\emph{Phys. Rev.} {\bfseries
  124} (1961) 1577}.

\bibitem{Arbuzov:2012nm}
A.B.~Arbuzov and T.V.~Kopylova, \emph{{Relativization of the
  Sommerfeld-Gamow-Sakharov factor}},
  \href{https://doi.org/10.1016/j.nuclphysbps.2012.02.006}{\emph{Nucl. Phys. B
  Proc. Suppl.} {\bfseries 225-227} (2012) 22}.

\bibitem{Belle:2007umv}
{\scshape Belle} collaboration, \emph{{Observation of Two Resonant Structures
  in $e^+e^- \to \pi^+ \pi^- \psi(2S)$ via Initial State Radiation at Belle}},
  \href{https://doi.org/10.1103/PhysRevLett.99.142002}{\emph{Phys. Rev. Lett.}
  {\bfseries 99} (2007) 142002}
  [\href{https://arxiv.org/abs/0707.3699}{{\ttfamily 0707.3699}}].

\bibitem{BESIII:2019nep}
{\scshape BESIII} collaboration, \emph{{Complete Measurement of the $\Lambda$
  Electromagnetic Form Factors}},
  \href{https://doi.org/10.1103/PhysRevLett.123.122003}{\emph{Phys. Rev. Lett.}
  {\bfseries 123} (2019) 122003}
  [\href{https://arxiv.org/abs/1903.09421}{{\ttfamily 1903.09421}}].

\bibitem{BESIII:2019odb}
{\scshape BESIII} collaboration, \emph{{Measurements of Weak Decay Asymmetries
  of $\Lambda_c^+\to pK_S^0$, $\Lambda\pi^+$, $\Sigma^+\pi^0$, and
  $\Sigma^0\pi^+$}},
  \href{https://doi.org/10.1103/PhysRevD.100.072004}{\emph{Phys. Rev. D}
  {\bfseries 100} (2019) 072004}
  [\href{https://arxiv.org/abs/1905.04707}{{\ttfamily 1905.04707}}].

\bibitem{Arleo:2008dn}
F.~Arleo, \emph{{(Medium-modified) Fragmentation Functions}},
  \href{https://doi.org/10.1140/epjc/s10052-009-0871-z}{\emph{Eur. Phys. J. C}
  {\bfseries 61} (2009) 603} [\href{https://arxiv.org/abs/0810.1193}{{\ttfamily
  0810.1193}}].

\bibitem{Collins:1992kk}
J.C.~Collins, \emph{{Fragmentation of transversely polarized quarks probed in
  transverse momentum distributions}},
  \href{https://doi.org/10.1016/0550-3213(93)90262-N}{\emph{Nucl. Phys. B}
  {\bfseries 396} (1993) 161}
  [\href{https://arxiv.org/abs/hep-ph/9208213}{{\ttfamily hep-ph/9208213}}].

\bibitem{BESIII:2015fyw}
{\scshape BESIII} collaboration, \emph{{Measurement of azimuthal asymmetries in
  inclusive charged dipion production in $e^+e^-$ annihilations at $\sqrt{s}$ =
  3.65 GeV}}, \href{https://doi.org/10.1103/PhysRevLett.116.042001}{\emph{Phys.
  Rev. Lett.} {\bfseries 116} (2016) 042001}
  [\href{https://arxiv.org/abs/1507.06824}{{\ttfamily 1507.06824}}].

\bibitem{BESIII:2018wid}
{\scshape BESIII} collaboration, \emph{{Measurement of the phase between strong
  and electromagnetic amplitudes of $J/\psi$ decays}},
  \href{https://doi.org/10.1016/j.physletb.2019.03.001}{\emph{Phys. Lett. B}
  {\bfseries 791} (2019) 375}
  [\href{https://arxiv.org/abs/1808.02166}{{\ttfamily 1808.02166}}].

\bibitem{BaldiniFerroli:2016mbs}
R.~Baldini~Ferroli, A.~Mangoni and S.~Pacetti, \emph{{$G$-parity violating
  amplitudes in the $J/\psi \to \pi^+ \pi^-$ decay}},
  \href{https://doi.org/10.1103/PhysRevC.98.045210}{\emph{Phys. Rev. C}
  {\bfseries 98} (2018) 045210}
  [\href{https://arxiv.org/abs/1611.04437}{{\ttfamily 1611.04437}}].

\bibitem{Perl:1975bf}
M.L.~Perl et~al., \emph{{Evidence for Anomalous Lepton Production in $e^+e^-$
  Annihilation}},
  \href{https://doi.org/10.1103/PhysRevLett.35.1489}{\emph{Phys. Rev. Lett.}
  {\bfseries 35} (1975) 1489}.

\bibitem{Asner:2009zza}
D.M.~Asner et~al., \emph{{Charm physics}},
  \href{https://doi.org/10.1142/S0217751X09046801}{\emph{Int. J. Mod. Phys. A}
  {\bfseries 24S1} (2009) 499}.

\bibitem{Abakumova:2011rp}
E.V.~Abakumova et~al., \emph{{The Beam Energy Measurement System for the
  Beijing Electron-Positron Collider}},
  \href{https://doi.org/10.1016/j.nima.2011.08.050}{\emph{Nucl. Instrum. Meth.
  A} {\bfseries 659} (2011) 21}
  [\href{https://arxiv.org/abs/1109.5771}{{\ttfamily 1109.5771}}].

\bibitem{Dai:2018thd}
L.R.~Dai, R.~Pavao, S.~Sakai and E.~Oset, \emph{{$\tau^- \to \nu_{\tau} M_1
  M_2$, with $M_1, M_2$ pseudoscalar or vector mesons}},
  \href{https://doi.org/10.1140/epja/i2019-12690-9}{\emph{Eur. Phys. J. A}
  {\bfseries 55} (2019) 20} [\href{https://arxiv.org/abs/1805.04573}{{\ttfamily
  1805.04573}}].

\bibitem{Sanchis-Lozano:1993vyw}
M.A.~Sanchis-Lozano, \emph{{On the search for weak decays of heavy quarkonium
  in dedicated heavy quark factories}},
  \href{https://doi.org/10.1007/BF01560243}{\emph{Z. Phys. C} {\bfseries 62}
  (1994) 271}.

\bibitem{Datta:1998yq}
A.~Datta, P.J.~O'Donnell, S.~Pakvasa and X.~Zhang, \emph{{Flavor changing
  processes in quarkonium decays}},
  \href{https://doi.org/10.1103/PhysRevD.60.014011}{\emph{Phys. Rev. D}
  {\bfseries 60} (1999) 014011}
  [\href{https://arxiv.org/abs/hep-ph/9812325}{{\ttfamily hep-ph/9812325}}].

\bibitem{Li:2012vk}
H.-B.~Li and S.-H.~Zhu, \emph{{Mini-review of rare charmonium decays at
  BESIII}}, \href{https://doi.org/10.1088/1674-1137/36/10/003}{\emph{Chin.
  Phys. C} {\bfseries 36} (2012) 932}
  [\href{https://arxiv.org/abs/1202.2955}{{\ttfamily 1202.2955}}].

\bibitem{Hill:1994hp}
C.T.~Hill, \emph{{Topcolor assisted technicolor}},
  \href{https://doi.org/10.1016/0370-2693(94)01660-5}{\emph{Phys. Lett. B}
  {\bfseries 345} (1995) 483}
  [\href{https://arxiv.org/abs/hep-ph/9411426}{{\ttfamily hep-ph/9411426}}].

\bibitem{BESIII:2014xbo}
{\scshape BESIII} collaboration, \emph{{Search for the rare decays $J/\psi \to
  D_s^-\rho^+$ and $J/\psi \to \overline{D}{}^0\overline{K}{}^{\ast0}$}},
  \href{https://doi.org/10.1103/PhysRevD.89.071101}{\emph{Phys. Rev. D}
  {\bfseries 89} (2014) 071101}
  [\href{https://arxiv.org/abs/1402.4025}{{\ttfamily 1402.4025}}].

\bibitem{BESIII:2014pps}
{\scshape BESIII} collaboration, \emph{{Search for the weak decays $J/\psi \to
  D^{(*)}_{s}e\nu_{e}+c.c.$}},
  \href{https://doi.org/10.1103/PhysRevD.90.112014}{\emph{Phys. Rev. D}
  {\bfseries 90} (2014) 112014}
  [\href{https://arxiv.org/abs/1410.8426}{{\ttfamily 1410.8426}}].

\bibitem{BESIII:2017pez}
{\scshape BESIII} collaboration, \emph{{Search for the rare decays $J/\psi \to
  D^{0} e^{+}e^{-} +c.c.$ and $\psi(3686) \to D^{0} e^{+}e^{-} +c.c.$}},
  \href{https://doi.org/10.1103/PhysRevD.96.111101}{\emph{Phys. Rev. D}
  {\bfseries 96} (2017) 111101}
  [\href{https://arxiv.org/abs/1710.02278}{{\ttfamily 1710.02278}}].

\bibitem{Bigi:2017eni}
I.I.~Bigi, X.-W.~Kang and H.-B.~Li, \emph{{CP Asymmetries in Strange Baryon
  Decays}}, \href{https://doi.org/10.1088/1674-1137/42/1/013101}{\emph{Chin.
  Phys. C} {\bfseries 42} (2018) 013101}
  [\href{https://arxiv.org/abs/1704.04708}{{\ttfamily 1704.04708}}].

\bibitem{BaBar:2012obs}
{\scshape BaBar} collaboration, \emph{{Evidence for an excess of $\bar{B} \to
  D^{(*)} \tau^-\bar{\nu}_\tau$ decays}},
  \href{https://doi.org/10.1103/PhysRevLett.109.101802}{\emph{Phys. Rev. Lett.}
  {\bfseries 109} (2012) 101802}
  [\href{https://arxiv.org/abs/1205.5442}{{\ttfamily 1205.5442}}].

\bibitem{BaBar:2013mob}
{\scshape BaBar} collaboration, \emph{{Measurement of an Excess of $\bar{B} \to
  D^{(*)}\tau^- \bar{\nu}_\tau$ Decays and Implications for Charged Higgs
  Bosons}}, \href{https://doi.org/10.1103/PhysRevD.88.072012}{\emph{Phys. Rev.
  D} {\bfseries 88} (2013) 072012}
  [\href{https://arxiv.org/abs/1303.0571}{{\ttfamily 1303.0571}}].

\bibitem{Belle:2007qnm}
{\scshape Belle} collaboration, \emph{{Observation of $B^0 \to D^{*-} \tau^+
  \nu_\tau$ decay at Belle}},
  \href{https://doi.org/10.1103/PhysRevLett.99.191807}{\emph{Phys. Rev. Lett.}
  {\bfseries 99} (2007) 191807}
  [\href{https://arxiv.org/abs/0706.4429}{{\ttfamily 0706.4429}}].

\bibitem{Belle:2010tvu}
{\scshape Belle} collaboration, \emph{{Observation of $B^+ \to \bar{D}^{*0}
  \tau^+ \nu_\tau$ and Evidence for $B^+ \to \bar{D}^0 \tau^+ \nu_\tau$ at
  Belle}}, \href{https://doi.org/10.1103/PhysRevD.82.072005}{\emph{Phys. Rev.
  D} {\bfseries 82} (2010) 072005}
  [\href{https://arxiv.org/abs/1005.2302}{{\ttfamily 1005.2302}}].

\bibitem{LHCb:2015gmp}
{\scshape LHCb} collaboration, \emph{{Measurement of the ratio of branching
  fractions $\mathcal{B}(\bar{B}^0 \to
  D^{*+}\tau^{-}\bar{\nu}_{\tau})/\mathcal{B}(\bar{B}^0 \to
  D^{*+}\mu^{-}\bar{\nu}_{\mu})$}},
  \href{https://doi.org/10.1103/PhysRevLett.115.111803}{\emph{Phys. Rev. Lett.}
  {\bfseries 115} (2015) 111803}
  [\href{https://arxiv.org/abs/1506.08614}{{\ttfamily 1506.08614}}].

\bibitem{LHCb:2016aa}
{\scshape LHCb} collaboration, \emph{{Angular analysis of the $B^0 \to
  K^{*0}\mu^+\mu^-$ decay using 3 fb$^{-1}$ of integrated luminosity}},
  \href{https://doi.org/10.1007/JHEP02(2016)104}{\emph{Journal of High Energy
  Physics} {\bfseries 2016} (2016) 104}
  [\href{https://arxiv.org/abs/1512.04442}{{\ttfamily 1512.04442}}].

\bibitem{LHCb:2015aa}
{\scshape LHCb} collaboration, \emph{{Measurement of the exclusive Y production
  cross-section in pp collisions at $\sqrt{s}=7$ TeV and 8 TeV}},
  \href{https://doi.org/10.1007/JHEP09(2015)084}{\emph{Journal of High Energy
  Physics} {\bfseries 2015} (2015) 84}
  [\href{https://arxiv.org/abs/1505.08139}{{\ttfamily 1505.08139}}].

\bibitem{LHCb:2014vgu}
{\scshape LHCb} collaboration, \emph{{Test of lepton universality using
  $B^{+}\rightarrow K^{+}\ell^{+}\ell^{-}$ decays}},
  \href{https://doi.org/10.1103/PhysRevLett.113.151601}{\emph{Phys. Rev. Lett.}
  {\bfseries 113} (2014) 151601}
  [\href{https://arxiv.org/abs/1406.6482}{{\ttfamily 1406.6482}}].

\bibitem{Nomura:2014asa}
Y.~Nomura and S.~Shirai, \emph{{Supersymmetry from Typicality: TeV-Scale
  Gauginos and PeV-Scale Squarks and Sleptons}},
  \href{https://doi.org/10.1103/PhysRevLett.113.111801}{\emph{Phys. Rev. Lett.}
  {\bfseries 113} (2014) 111801}
  [\href{https://arxiv.org/abs/1407.3785}{{\ttfamily 1407.3785}}].

\bibitem{Altmannshofer:2017yso}
W.~Altmannshofer, P.~Stangl and D.M.~Straub, \emph{{Interpreting Hints for
  Lepton Flavor Universality Violation}},
  \href{https://doi.org/10.1103/PhysRevD.96.055008}{\emph{Phys. Rev. D}
  {\bfseries 96} (2017) 055008}
  [\href{https://arxiv.org/abs/1704.05435}{{\ttfamily 1704.05435}}].

\bibitem{Dorsner2017}
I.~Doršner, S.~Fajfer, D.A.~Faroughy and N.~Košnik, \emph{{The role of the S3
  GUT leptoquark in flavor universality and collider searches}},
  \href{https://doi.org/10.1007/JHEP10(2017)188}{\emph{Journal of High Energy
  Physics} {\bfseries 2017} (2017) 188}
  [\href{https://arxiv.org/abs/1706.07779}{{\ttfamily 1706.07779}}].

\bibitem{Fajfer:2015ixa}
S.~Fajfer, I.~Nisandzic and U.~Rojec, \emph{{Discerning new physics in charm
  meson leptonic and semileptonic decays}},
  \href{https://doi.org/10.1103/PhysRevD.91.094009}{\emph{Phys. Rev. D}
  {\bfseries 91} (2015) 094009}
  [\href{https://arxiv.org/abs/1502.07488}{{\ttfamily 1502.07488}}].

\bibitem{BESIII:2018cls}
{\scshape BESIII} collaboration, \emph{{Search for Baryon and Lepton Number
  Violation in $J/\psi\to\Lambda_c^+e^-+c.c.$}},
  \href{https://doi.org/10.1103/PhysRevD.99.072006}{\emph{Phys. Rev. D}
  {\bfseries 99} (2019) 072006}
  [\href{https://arxiv.org/abs/1803.04789}{{\ttfamily 1803.04789}}].

\bibitem{BESIII:2019oef}
{\scshape BESIII} collaboration, \emph{{Search for heavy Majorana neutrino in
  lepton number violating decays of $D\to K \pi e^+ e^+$}},
  \href{https://doi.org/10.1103/PhysRevD.99.112002}{\emph{Phys. Rev. D}
  {\bfseries 99} (2019) 112002}
  [\href{https://arxiv.org/abs/1902.02450}{{\ttfamily 1902.02450}}].

\bibitem{BESIII:2019udi}
{\scshape BESIII} collaboration, \emph{{Search for baryon and lepton number
  violating decays $D^+\to\bar\Lambda(\bar\Sigma^0)e^+$ and
  $D^+\to\Lambda(\Sigma^0)e^+$}},
  \href{https://doi.org/10.1103/PhysRevD.101.031102}{\emph{Phys. Rev. D}
  {\bfseries 101} (2020) 031102}
  [\href{https://arxiv.org/abs/1911.13116}{{\ttfamily 1911.13116}}].

\bibitem{BESIII:2021krj}
{\scshape BESIII} collaboration, \emph{{Search for baryon- and lepton-number
  violating decays $D^0 \rightarrow \overline{p} e^+$ and $D^0 \rightarrow
  pe^-$}}, \href{https://doi.org/10.1103/PhysRevD.105.032006}{\emph{Phys. Rev.
  D} {\bfseries 105} (2022) 032006}
  [\href{https://arxiv.org/abs/2112.10972}{{\ttfamily 2112.10972}}].

\bibitem{BESIII:2020iwk}
{\scshape BESIII} collaboration, \emph{{Search for the lepton number violating
  decay $\Sigma^{-} \to p e^{-} e^{-}$ and the rare inclusive decay $\Sigma^{-}
  \to \Sigma^{+} X$}},
  \href{https://doi.org/10.1103/PhysRevD.103.052011}{\emph{Phys. Rev. D}
  {\bfseries 103} (2021) 052011}
  [\href{https://arxiv.org/abs/2012.03592}{{\ttfamily 2012.03592}}].

\bibitem{Li:2016tlt}
H.-B.~Li, \emph{{Prospects for rare and forbidden hyperon decays at BESIII}},
  \href{https://doi.org/10.1007/s11467-017-0691-9}{\emph{Front. Phys.
  (Beijing)} {\bfseries 12} (2017) 121301}
  [\href{https://arxiv.org/abs/1612.01775}{{\ttfamily 1612.01775}}].

\bibitem{BESIII:2021slv}
{\scshape BESIII} collaboration, \emph{{Search for Invisible Decays of the
  $\Lambda$ Baryon}},  \href{https://arxiv.org/abs/2110.06759}{{\ttfamily
  2110.06759}}.

\bibitem{Mohapatra:1980qe}
R.N.~Mohapatra and R.E.~Marshak, \emph{{Local B-L Symmetry of Electroweak
  Interactions, Majorana Neutrinos and Neutron Oscillations}},
  \href{https://doi.org/10.1103/PhysRevLett.44.1316}{\emph{Phys. Rev. Lett.}
  {\bfseries 44} (1980) 1316}.

\bibitem{Kang:2009xt}
X.-W.~Kang, H.-B.~Li and G.-R.~Lu, \emph{{Study of $\Lambda - \bar{\Lambda}$
  Oscillation in quantum coherent $\Lambda \bar{\Lambda}$ state by using
  $J/\psi \to \Lambda \bar{\Lambda}$ decay}},
  \href{https://doi.org/10.1103/PhysRevD.81.051901}{\emph{Phys. Rev. D}
  {\bfseries 81} (2010) 051901}
  [\href{https://arxiv.org/abs/0906.0230}{{\ttfamily 0906.0230}}].

\bibitem{BES:2003zru}
{\scshape BES} collaboration, \emph{{Search for lepton flavor violation process
  $J / \psi \to e \mu$}},
  \href{https://doi.org/10.1016/S0370-2693(03)00391-5}{\emph{Phys. Lett. B}
  {\bfseries 561} (2003) 49}
  [\href{https://arxiv.org/abs/hep-ex/0303005}{{\ttfamily hep-ex/0303005}}].

\bibitem{BES:2004jiw}
{\scshape BES} collaboration, \emph{{Search for the lepton flavor violation
  processes $J / \psi \to \mu \tau$ and $e \tau$}},
  \href{https://doi.org/10.1016/j.physletb.2004.08.005}{\emph{Phys. Lett. B}
  {\bfseries 598} (2004) 172}
  [\href{https://arxiv.org/abs/hep-ex/0406018}{{\ttfamily hep-ex/0406018}}].

\bibitem{BESIII:2013jau}
{\scshape BESIII} collaboration, \emph{{Search for the lepton flavor violation
  process J/\ensuremath{\psi}\textrightarrow{}e\ensuremath{\mu} at BESIII}},
  \href{https://doi.org/10.1103/PhysRevD.87.112007}{\emph{Phys. Rev. D}
  {\bfseries 87} (2013) 112007}
  [\href{https://arxiv.org/abs/1304.3205}{{\ttfamily 1304.3205}}].

\bibitem{LHCb:2015pce}
{\scshape LHCb} collaboration, \emph{{Search for the lepton-flavour violating
  decay $D^0 \to e^\pm\mu^\mp$}},
  \href{https://doi.org/10.1016/j.physletb.2016.01.029}{\emph{Phys. Lett. B}
  {\bfseries 754} (2016) 167}
  [\href{https://arxiv.org/abs/1512.00322}{{\ttfamily 1512.00322}}].

\bibitem{Liu:2018jdi}
Z.~Liu and Y.~Zhang, \emph{{Probing millicharge at BESIII via monophoton
  searches}}, \href{https://doi.org/10.1103/PhysRevD.99.015004}{\emph{Phys.
  Rev. D} {\bfseries 99} (2019) 015004}
  [\href{https://arxiv.org/abs/1808.00983}{{\ttfamily 1808.00983}}].

\bibitem{Liang:2019zkb}
J.~Liang, Z.~Liu, Y.~Ma and Y.~Zhang, \emph{{Millicharged particles at electron
  colliders}}, \href{https://doi.org/10.1103/PhysRevD.102.015002}{\emph{Phys.
  Rev. D} {\bfseries 102} (2020) 015002}
  [\href{https://arxiv.org/abs/1909.06847}{{\ttfamily 1909.06847}}].

\bibitem{Krasznahorkay:2015iga}
A.J.~Krasznahorkay et~al., \emph{{Observation of Anomalous Internal Pair
  Creation in Be8 : A Possible Indication of a Light, Neutral Boson}},
  \href{https://doi.org/10.1103/PhysRevLett.116.042501}{\emph{Phys. Rev. Lett.}
  {\bfseries 116} (2016) 042501}
  [\href{https://arxiv.org/abs/1504.01527}{{\ttfamily 1504.01527}}].

\bibitem{Krasznahorkay:2019lyl}
A.J.~Krasznahorkay et~al., \emph{{New evidence supporting the existence of the
  hypothetic X17 particle}},
  \href{https://arxiv.org/abs/1910.10459}{{\ttfamily 1910.10459}}.

\bibitem{Jiang:2018jqp}
J.~Jiang, H.~Yang and C.-F.~Qiao, \emph{{Exploring Bosonic Mediator of
  Interaction at BESIII}},
  \href{https://doi.org/10.1140/epjc/s10052-019-6912-3}{\emph{Eur. Phys. J. C}
  {\bfseries 79} (2019) 404}
  [\href{https://arxiv.org/abs/1810.05790}{{\ttfamily 1810.05790}}].

\bibitem{Belle-II:2020jti}
{\scshape Belle-II} collaboration, \emph{{Search for Axion-Like Particles
  produced in $e^+e^-$ collisions at Belle II}},
  \href{https://doi.org/10.1103/PhysRevLett.125.161806}{\emph{Phys. Rev. Lett.}
  {\bfseries 125} (2020) 161806}
  [\href{https://arxiv.org/abs/2007.13071}{{\ttfamily 2007.13071}}].

\bibitem{BESIII:2017fwv}
{\scshape BESIII} collaboration, \emph{{Dark Photon Search in the Mass Range
  Between 1.5 and 3.4 GeV/$c^2$}},
  \href{https://doi.org/10.1016/j.physletb.2017.09.067}{\emph{Phys. Lett. B}
  {\bfseries 774} (2017) 252}
  [\href{https://arxiv.org/abs/1705.04265}{{\ttfamily 1705.04265}}].

\bibitem{BaBar:2014zli}
{\scshape BaBar} collaboration, \emph{{Search for a Dark Photon in $e^+e^-$
  Collisions at BaBar}},
  \href{https://doi.org/10.1103/PhysRevLett.113.201801}{\emph{Phys. Rev. Lett.}
  {\bfseries 113} (2014) 201801}
  [\href{https://arxiv.org/abs/1406.2980}{{\ttfamily 1406.2980}}].

\bibitem{Alonso-Alvarez:2021oaj}
G.~Alonso-\'Alvarez, G.~Elor, M.~Escudero, B.~Fornal, B.~Grinstein and
  J.M.~Camalich, \emph{{The Strange Physics of Dark Baryons}},
  \href{https://arxiv.org/abs/2111.12712}{{\ttfamily 2111.12712}}.

\end{thebibliography}\endgroup
\bibliographystyle{JHEP}

\end{document}